\documentclass[preprint,emulateapj,twocolumn]{aastex}
\usepackage{epsf}
\usepackage{graphicx}
\usepackage{emulateapj5}
\usepackage{onecolfloat}
\usepackage{apjfonts}
\usepackage{amsmath}

\newcommand{\tha}{$^{\mathrm{th}}$}

\newcommand{\lsim}{\lower0.6ex\vbox{\hbox{$ \buildrel{\textstyle <}\over{\sim}\ $}}}
\newcommand{\gsim}{\lower0.6ex\vbox{\hbox{$ \buildrel{\textstyle >}\over{\sim}\ $}}}

\newcommand{\hkpc}{h^{-1}\mathrm{kpc}}
\newcommand{\hpc}{h^{-1}\mathrm{pc}}
\newcommand{\hMsun}{\ h^{-1}\mathrm{M}_{\odot}}
\newcommand{\hMpc}{\ h^{-1}\mathrm{Mpc}}

\newcommand{\kms}{{\,{\rm km}\,{\rm s}^{-1}}}
\newcommand{\kpc}{{\,{\rm kpc}}}
\newcommand{\Gyr}{{\,{\rm Gyr}}}
\newcommand{\mpt}{m_{\mathrm{p}}}

\newcommand{\feh}{[\mathrm{Fe}/\mathrm{H}]}

\newcommand{\Tmw}{\Theta_{\mathrm{MW}}}
\newcommand{\Tsats}{\Theta_{\mathrm{SATS}}}
\newcommand{\degrees}{^{\circ}}

\newcommand{\Omegam}{\Omega_{M}}

\newcommand{\Omegal}{\Omega_{\Lambda}}

\newcommand{\sig}{\sigma_{8}}
\newcommand{\rhomean}{\rho_{\mathrm{M}}}

\newcommand{\vmax}{V_{\mathrm{max}}}

\newcommand{\Rvir}{R_{180}}
\newcommand{\Mvir}{M_{180}}

\newcommand{\Vcirc}{V_{\mathrm{circ}}}
\newcommand{\Vmax}{V_{\mathrm{max}}}
\newcommand{\rs}{r_{\mathrm{s}}}

\newcommand{\cvir}{c_{\mathrm{180}}}

\newcommand{\rfind}{r_{\mathrm{find}}}
\newcommand{\rt}{r_{\mathrm{t}}}
\newcommand{\rp}{r_{\mathrm{p}}}
\newcommand{\Rcut}{r_{\mathrm{cut}}}

\newcommand{\tdyn}{t_{\mathrm{dyn}}}

\newcommand{\gone}{\mathrm{G}_{1}}
\newcommand{\gtwo}{\mathrm{G}_{2}}
\newcommand{\gthree}{\mathrm{G}_{3}}

\newcommand{\Pks}{P_{\mathrm{KS}}}
\newcommand{\Pksiso}{P_{\mathrm{KS}}^{\mathrm{ISO}}}

\newcommand{\Pkst}{P_{\mathrm{KS}}^{\theta}}

\newcommand{\fsat}{f_{\mathrm{sat}}}

\newcommand{\acost}{\vert \cos(\theta)\vert}
\newcommand{\cosw}{\cos(\omega)}
\newcommand{\acosw}{\vert \cos(\omega)\vert}

\newcommand{\drms}{D_{\mathrm{rms}}}
\newcommand{\delrms}{\delta_{\mathrm{rms}}}
\newcommand{\drmsiso}{D_{\mathrm{rms}}^{\mathrm{iso}}}
\newcommand{\drmsmw}{D_{\mathrm{rms}}^{\mathrm{MW}}}
\newcommand{\rmed}{R_{\mathrm{med}}}

\newcommand{\beq}{\begin{equation}}
\newcommand{\eeq}{\end{equation}}

\newcommand{\dd}{\mathrm{d}}

\begin{document}

\submitted{The Astrophysical Journal, submitted}
\vspace{1mm}
\slugcomment{{\em The Astrophysical Journal, submitted}} 

\shortauthors{Zentner et al.}

\twocolumn[
\lefthead{Anisotropic Distribution of Galactic Satellites}
\righthead{Zentner et al.}

\title{The Anisotropic Distribution of Galactic Satellites}

\author{
Andrew R. Zentner\altaffilmark{1}, 
Andrey V. Kravtsov\altaffilmark{1,2},
Oleg Y. Gnedin\altaffilmark{3},
Anatoly A. Klypin\altaffilmark{4}
}

\begin{abstract}
  We present a study of the spatial distribution of dwarf satellites
  (or subhalos) in galactic dark matter halos using dissipationless
  cosmological simulations of the concordance flat Cold Dark Matter
  (CDM) model with vacuum energy.  We find that subhalos are
  distributed anisotropically and are preferentially located along the
  major axes of the triaxial mass distributions of their hosts.  The
  Kolmogorov-Smirnov probability for drawing our simulated subhalo
  sample from an isotropic distribution is $\Pks \simeq 1.5 \times
  10^{-4}$.  An isotropic distribution of subhalos is thus not the
  correct null hypothesis for testing the CDM paradigm.  The nearly
  planar distribution of observed Milky Way (MW) satellites is
  marginally consistent (probability $\simeq 0.02$) with being
  drawn randomly from the subhalo distribution in our simulations.
  Furthermore, if we select the subhalos likely to be luminous, 
  we find a distribution that is consistent with the observed 
  MW satellites.  In fact, we show that subsamples of the 
  subhalo population with a centrally-concentrated radial 
  distribution that is similar to that of the MW dwarfs typically 
  exhibit a comparable degree of planarity.  We explore 
  the origin of the observed subhalo anisotropy and conclude 
  that it is likely due to (1) the preferential accretion of satellites 
  along filaments, often closely aligned with the major axis of the 
  host halo, and (2) evolution of satellite orbits within the prolate,
  triaxial potentials typical of CDM halos. Agreement between
  predictions and observations requires the major axis of the outer
  dark matter halo of the Milky Way to be nearly perpendicular to the
  disk. We discuss possible observational tests of such disk--halo
  alignment with current large galaxy surveys.
\end{abstract}

\keywords{cosmology: theory, large-scale structure of universe 
-- dark matter -- 
galaxies: formation, halos, structure --- methods: numerical}
]


\altaffiltext{1}{
Kavli Institute for Cosmological Physics and
Department of Astronomy and Astrophysics, 
The University of Chicago, 
933 East 56\tha Street,
Chicago, IL 60637 USA;  
{\tt zentner@kicp.uchicago.edu,andrey@oddjob.uchicago.edu}
}

\altaffiltext{2}{
The Enrico Fermi Institute, 
5640 South Ellis Ave.,
The University of Chicago, 
Chicago, IL 60637 USA
}
\altaffiltext{3}{
Department of Astronomy, 
The Ohio State University, 
140 W. 18\tha Ave., 
Columbus, OH 43210 USA; 
ognedin@astronomy.ohio-state.edu
}
\altaffiltext{4}{
Astronomy Department, 
New Mexico State University, 
MSC 4500, P.O. Box 30001, Las Cruces, NM 880003-8001; 
{\tt aklypin@nmsu.edu}
}


%
%
\section{Introduction}
\label{sec:intro}

Simulations of structure formation in the standard cold dark matter
(CDM) scenario \citep[e.g.,][]{blumenthal_etal84} show that virialized 
dark matter halos teem with distinct, gravitationally-bound 
substructures, often referred to as {\em subhalos}.  
The abundance of subhalos in Milky Way-sized 
halos has received much attention, as there are
more than an order of magnitude fewer observed dwarf satellite
galaxies around these systems than the predicted number of subhalos of
comparable velocity dispersion \citep{kauffmann_etal93,
  klypin_etal99a, moore_etal99}.  This problem 
is often referred to as the ``the missing satellites problem.''

The difference in predicted and observed abundances of dwarf
satellites is likely related to the corresponding difference in
spatial distributions. The predicted radial distribution of subhalo
populations is considerably more extended than that of the 
Milky Way (MW) satellites 
\citep{taylor_etal03, kravtsov_etal04}.  This may indicate
that the environments in which low-luminosity galaxies form are
non-trivially biased relative to the overall population of subhalos.
In addition, the dwarf satellites of the MW and M31 appear to be
distributed anisotropically about their hosts
\citep[e.g.,][]{lynden-bell82, majewski94, hartwick96, hartwick00,
  mateo98, grebel_etal99, willman_etal04}.  \citet{kroupa_etal04}
recently argued that the anisotropic distribution of the MW dwarf 
satellites presents a serious challenge to the standard CDM
structure formation paradigm.  \citet{kroupa_etal04} reached this
conclusion by assuming that the luminous satellites correspond to a
randomly-selected subset of dark matter subhalos and by taking an
isotropic subhalo distribution as their null hypothesis.

There have been a number of studies of anisotropy in satellite galaxy
distributions outside of the Local Group.  
\citet{holmberg69} found that satellites of spiral galaxies with
projected separations $\rp \lsim 50 \kpc$ are preferentially 
located near the short axes of the projected light distributions 
of their host galaxies (the {\em Holmberg Effect}).  
However, several subsequent studies found little evidence for such a
preferential alignment \citep[e.g.,][]{hawley_peebles75, sharp_etal79,
  macgillivray_etal82}.  \citet{zaritsky_etal97} found a statistically
significant anisotropy similar to that advocated by
\citet{holmberg69}, but only at larger projected distances, 
$200\, \kpc\, \lsim \rp \lsim\, 500\, \kpc$.  
In a more recent study of a
large sample of satellite galaxies in the Two Degree Field Galaxy
Redshift Survey \citep{colless_etal01}, \citet{sales_lambas04} also
found evidence for preferential alignment of satellites along the 
minor axis of the central galaxy.  
However, \citet{brainerd04} performed a similar study of
satellites in the Sloan Digital Sky Survey 
\citep{york_etal00,strauss_etal02}, finding satellites to be
preferentially aligned near the {\em long} axes of 
host galaxies and the degree of anisotropy 
to be a rapidly decreasing function of
separation from the host galaxy.  Note that the anisotropy measured by
\citet{brainerd04} is {\em opposite} in sense to the anisotropy 
reported by \citet{holmberg69}, \citet{zaritsky_etal97}, and
\citet{sales_lambas04}.

Properties of dwarf satellite dark matter (DM) halos 
(or {\em subhalos}) in MW-sized hosts have 
been the subject of several
recent studies, which used a new generation of high-resolution
dissipationless simulations not affected by the ``overmerging''
problem
\citep[e.g.,][]{klypin_etal99a,moore_etal99,stoehr_etal02,stoehr_etal03,
  delucia_etal04,kravtsov_etal04,gao_etal04a,reed_etal04}.  One of the
main results is that the radial distribution of subhalos is more
extended than the distribution of DM
\citep{ghigna_etal98,colin_etal99,ghigna_etal00,springel_etal01,
  delucia_etal04,gao_etal04b}. The DM subhalos also appear to
have a significantly more extended and shallower radial distribution
compared to the observed distribution of satellite galaxies both in
galactic halos \citep{taylor_etal03, kravtsov_etal04} and in cluster halos
\citep{diemand_etal04,gao_etal04b,nagai_kravtsov05}. Despite
significant progress in our understanding of halo substructure, the
anisotropy of subhalo spatial distributions has, so far, not been
studied in as much detail.  \citet{zaritsky_etal97} reported no detectable
anisotropy in the projected satellite distribution in the simulations
of \citet*{navarro_etal94,navarro_etal95}.  However, this conclusion
refers to the statistical, projected distribution of the most massive
satellites obtained after stacking many galaxy-sized halos, rather
than anisotropy of the satellite distribution within a single host
halo. More recently, \citet{knebe_etal04} studied anisotropy of the 
subhalo distribution in dissipationless simulations of cluster-sized
hosts. These authors found that subhalo distribution is anisotropic
and is aligned with the major axis of the matter distribution of the
host.

In this paper, we study the anisotropy of satellite distribution in
Galaxy-sized halos using high-resolution cosmological $N$-body
simulations of structure formation in the {\em concordance} flat
$\Lambda$CDM cosmology.  We show that an isotropic distribution is not
the correct null hypothesis for testing the CDM paradigm.  The mass
distributions in CDM halos are generally triaxial rather than
spherical.  We demonstrate that subhalos of the size thought to host
the MW dwarf satellites are distributed anisotropically about their
host halos with subhalos preferentially located along the major axes
of their hosts.  We also show that the null hypothesis distribution
taken by \citet{kroupa_etal04} is incorrect even in the case of an
isotropic underlying distribution.  As we were completing this study, 
\citet{kang_etal05} presented a similar study, 
considering in particular the question of whether the
anisotropy of the MW satellite distribution is consistent with the
hierarchical formation scenario.  Although our approaches differ in
detail, their conclusions are consistent with ours.  We present a more
extensive study of the satellite distribution, both in three
dimensions and in two-dimensional projection.  We also explore the
physical mechanisms that create the anisotropy measured in the
simulations and discuss the implications of our results on our
understanding of galaxy formation.

This manuscript is organized as follows.  In \S~\ref{sec:methods}, we
describe our numerical simulations and our analysis methods.  In
\S~\ref{sec:results}, we present results on the anisotropic
distribution of subhalos in Galaxy-sized halos.  We discuss our
results and their implications in \S~\ref{sec:disc}.  Lastly, we
summarize our main findings and draw conclusions in \S~\ref{sec:conc}.
In an Appendix, we discuss the prospects of studying 
satellite anisotropy in projection. 

Throughout this paper, we refer to halos that are contained 
within the virial radii of still larger halos as 
{\em subhalos} or {\em satellites} and we refer to halos 
that are not contained within a larger halo as 
{\em host} halos.

\section{Methods}
\label{sec:methods}

\subsection{Numerical Simulations}
\label{sub:sims}

We analyze a simulation of three Milky Way-sized DM halos
formed in a standard, ``concordance'' $\Lambda$CDM cosmology with
$\Omegam = 1 - \Omegal = 0.3$, $h = 0.7$, and $\sig = 0.9$.  The
simulation was performed with the Adaptive Refinement Tree (ART)
$N$-body code \citep{kravtsov_etal97,kravtsov99}.  This simulation has
been discussed previously by \citet{klypin_etal01}, and
\citet*[][hereafter KGK04]{kravtsov_etal04}.  We briefly review some
of the details in this section.

The simulation began with a uniform $256^3$ grid covering a comoving,
cubic box of $25\hMpc$ on a side.  Higher force resolution was
achieved in dense regions using an adaptive refinement algorithm.  The
grid cells were refined if the particle mass contained within them
exceeded a certain, specified threshold value.  The multiple mass
resolution technique was used to set up the initial conditions.  A
Lagrangian region corresponding to two virial radii about each halo
was re-sampled at the initial epoch of $z_i = 50$ with the
highest-resolution particles of mass $\mpt = 1.2 \times 10^{6}
\hMsun$.  The high mass resolution region was surrounded by layers of
particles of increasing mass with a total of five particle species.
Only the regions containing the highest-resolution particles were
adaptively refined.  The maximum level of refinement in the
simulations corresponded to a peak formal spatial resolution of
approximately $100 \hpc$.

We define a halo radius $\Rvir$, by the sphere, centered on the
particle with the highest density, within which the mean density is
$180$ times the mean density of the universe $\rhomean$, so that the
mass and radius are related by $\Mvir = 4\pi (180 \rhomean) \Rvir^3 /
3$.  Our three host halos, which we refer to as {\it halo} $\gone$,
{\it halo} $\gtwo$, and {\it halo} $\gthree$, have masses of $\Mvir =
1.66 \times 10^{12} \hMsun$, $1.24 \times 10^{12} \hMsun$, and
$1.19 \times 10^{12} \hMsun$, respectively, and these halos
contain $\sim 10^{6}$ particles within their virial radii.\footnote{
  These halos were referred to as $B_2$, $C_2$, and $D_2$ respectively
  by \citet{klypin_etal01}.}  Their virial radii are $\Rvir = 298
\hkpc$, $278 \hkpc$, and $281 \hkpc$, respectively.

We identified halos and subhalos using a variant of the Bound Density
Maxima algorithm \citep{klypin_etal99b}.  First, we compute the local
density at each particle using a smoothing kernel of $24$ particles
and identify local maxima in the density field.  Beginning with the
highest density particles and stepping down in density, we mark each
peak as a potential halo center and surround the peak by a sphere of
radius $\rfind = 10 \hkpc$.  All particles within the sphere are
excluded from further consideration as potential halo centers.  The
parameter $\rfind$ is set according to the smallest objects that we
aim to identify robustly.  After identifying potential halo centers,
we iteratively remove unbound particles.  For host halos, the mass and
radius are set according to a fixed overdensity as described above.
For subhalos, the outer boundary is somewhat ambiguous and we adopt a
truncation radius $\rt$, at which the density profile becomes greater
than a critical value of $\dd \ln \rho /\dd \ln r = -0.5$.  This
criterion is based on the fact that we do not expect density profiles
of CDM halos to be shallower than this and, empirically, this
definition is approximately equal to the radius at which the
background density of host halo particles is equal to the density of
particles bound to the subhalo.

Upon identifying halos, we assign each halo a mass and radius and use
the halo particles to determine a circular velocity profile 
$\Vcirc(r) = \sqrt{GM(<r)/r}$, 
and the maximum circular velocity, $\Vmax$.  We
choose to quantify the size of subhalos according to 
$\Vmax$ because this quantity is measured more robustly 
and is not subject to the same ambiguity as a 
particular mass definition.

Halos $\gone$, $\gtwo$, and $\gthree$ have maximum circular velocities
of $\Vmax \simeq 213 \kms$, $\Vmax \simeq 199 \kms$, and $\Vmax \simeq
183 \kms$ respectively. These halos were selected to reside in a
well-defined filament at $z=0$.  The halos $\gtwo$ and $\gthree$ are
neighbors located at $425h^{-1}\ \rm kpc$ 
(i.e., $\approx 610\ {\rm  kpc}\sim 2R_{\rm vir}$) 
from each other.  The configuration of this
pair thus resembles that of the Local Group. The third halo is
isolated and is located $\sim 2$~Mpc away from the pair.  The three
hosts have similar masses at the present but rather different mass
accretion histories.  Host $\gone$ undergoes a spectacular multiple
major merger at $z\approx 2$, which results in a dramatic mass
increase on a dynamical time scale.  Halos $\gtwo$ and $\gthree$ increase
their mass in a series of somewhat less spectacular major mergers
which could be seen as mass jumps at $5 < z < 1$.  All three systems
accrete little mass and experience no major mergers at $z\lesssim 1$
(or lookback time of $\approx 8$~Gyr) and thus could host a disk
galaxy. Note, however, that halos $\gone$ and $\gthree$ experience minor
mergers during this period.

\subsection{Modeling Luminous Satellites}
\label{sub:lumsats}

The relative scarcity of MW satellites compared 
to predicted subhalo counts suggests that a 
na{\"{\i}}ve comparison of subhalo populations to 
luminous galaxies may not be correct.  
This implies that in order to make a more 
meaningful comparison with the data, we should have 
a model to identify the DM subhalos in 
simulations that would host observable, 
luminous galaxies. 
We consider two alternatives.

The first dwarf galaxy formation model we consider was 
recently proposed by KGK04 (see their \S~6 for details).  
This semi-analytic model is based on the subhalo 
evolutionary tracks extracted from the 
simulations used in this study.  
The small-mass dwarfs are identified with the halos 
that either have accreted a large fraction of their
mass prior to the epoch of reionization 
\citep*[see also][]{bullock_etal00,ricotti_gnedin04} 
or were relatively massive at high redshifts and could 
therefore retain most of their gas and form stars after 
reionization.  Some objects in the latter 
category could lose most of their former mass 
due to tidal stripping and appear as relatively 
low-mass halos at the present epoch.  The dwarf 
satellite galaxies in this model can 
thus be hosted by both massive and 
low-mass subhalos at $z=0$. 

The physical ingredients of the KGK04 model include: 
(1) the suppression of gas accretion by the 
extragalactic UV background; 
(2) an observationally-motivated recipe 
for quiescent star formation; 
(3) a starburst mode of star formation 
after strong peaks in tidal forces 
(which are calculated self-consistently from the simulations); 
and (4) an accounting for the inefficient dissipation of 
gas in halos with $T_{\mathrm{vir}} \lesssim 10^4$~K.  
The model can successfully reproduce 
the circular velocity function, 
the radial distribution, 
and the morphological segregation of the 
observed MW satellites, 
as well as the basic
properties of galactic dwarfs such as 
their star formation histories, 
stellar masses, and densities.  
In our analysis, we use the same set of
subhalos that were considered to be ``luminous'' 
according to this model in KGK04 and refer to them 
as {\em luminous subhalos}.

The second model assumes that the observed MW satellites 
are hosted by the most massive subhalos
\citep{stoehr_etal02,stoehr_etal03}.  
In this model, it is conjectured that the masses of 
the subhalos in which the luminous dwarfs are embedded are 
significantly underestimated because the DM 
density profiles in the central regions of the 
subhalos have been affected by tidal interactions 
\citep[][see, however, \citeauthor{kazantzidis_etal04b}
\citeyear{kazantzidis_etal04b}]{stoehr_etal02,stoehr_etal03,
  hayashi_etal03}. 
The maximum circular velocities of the Local Group dwarfs 
may thus be systematically underestimated in observations 
because they are derived from stellar velocity dispersion 
measurements within radii considerably smaller than 
the radius at which $\Vmax$ is achieved.  
\citet{stoehr_etal02} argued that the bias induced by 
this is large, such that all of the observed MW 
satellites can be embedded in the most massive subhalos 
with $\Vmax \gsim 30-40 \kms$. This model has an 
important physical implication.  If the MW dwarfs inhabit 
the most massive subhalos, then there must exist a certain 
universal mass or $\Vmax$ scale below which 
galaxy formation is {\em completely} quashed due 
to the UV background heating and inefficient gas 
cooling in dwarf halos.  
We consider this type of model by associating 
luminous dwarfs with the eleven subhalos with the 
highest values of $\Vmax$ at $z=0$ within 
$300 \kpc$ of each MW-sized host halo.

\subsection{The Principal Axes of the Host Halos}
\label{sub:inertia}

\begin{figure}[t]
\epsscale{0.80}
\plotone{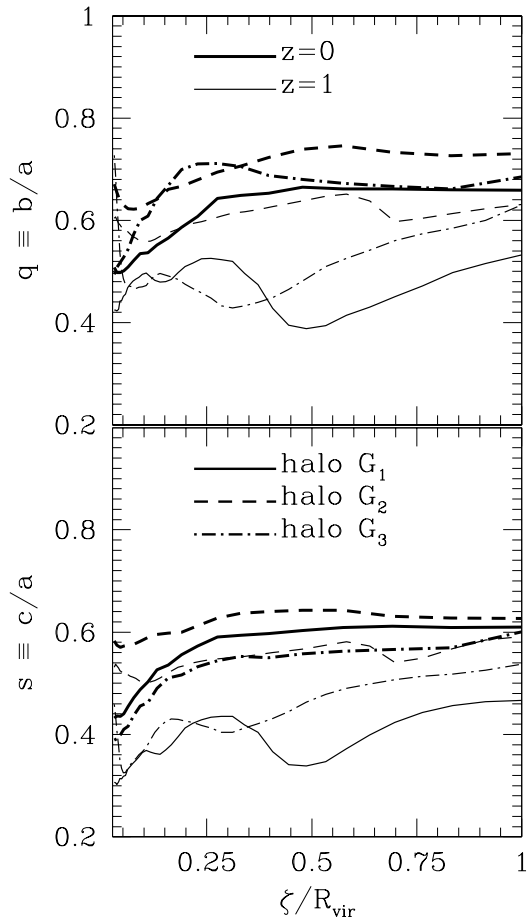}
\caption{
Host halo axis ratio profiles for the three MW-sized host halos.  
{\it Top}:  intermediate-to-long axis ratio $q \equiv b/a$, 
as a function of long axis length.
{\it Bottom}:  short-to-long axis ratio $s \equiv c/a$.  
In both panels, the 
{\it solid} line represents halo $\gone$, 
the {\it dashed} line represents halo $\gtwo$, 
and the {\it dot-dashed} line represents halo $\gthree$.  
The {\em thick} lines represent the shape 
profiles at $z=0$, 
while the {\em thin} lines represent 
the halo shape profiles at $z=1$.
}
\label{fig:shapes}
\end{figure}

\begin{figure*}[th]
\epsscale{2.0}
\plotone{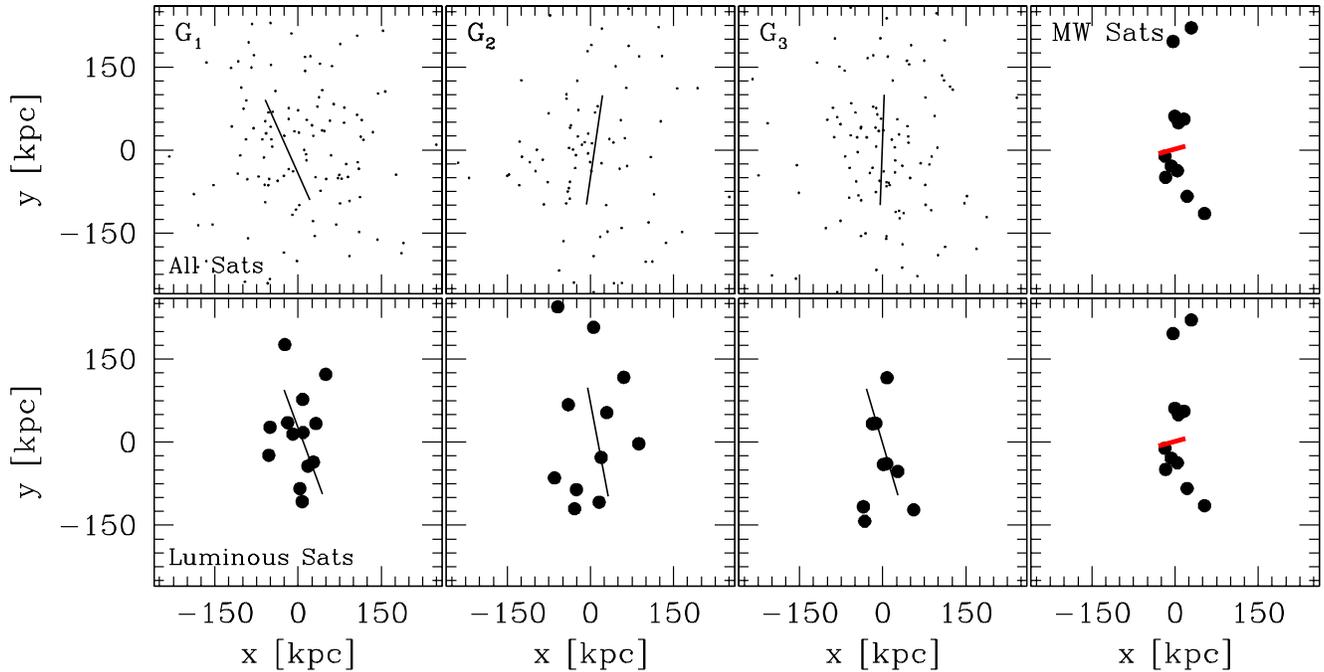}
\caption{
Projections of satellites in a plane orthogonal to their 
best-fit planes (see text).  All panels show scatter plots 
of the positions of satellites projected onto a plane perpendicular 
to their best-fit plane.  The best-fit plane corresponds 
to the vertical axis in this projection.  
In both rows, the first panel shows results 
for the subhalos of halo $\gone$, the second 
column for halo $\gtwo$, and the third column for halo 
$\gthree$.  The fourth column in each row shows the observed MW 
satellites.  In the first three columns, the projection is such 
that the major axis of the host halo lies in the plane of the 
projection.  In these panels, the major axis is shown as the 
{\em thin, solid} line.  In the fourth column, the projection is 
such that the MW disk is seen edge-on and the MW disk orientation 
is denoted by the {\em thick, solid} line.  
In the top row, we compute the best-fit plane by considering 
all subhalos within $\approx 300$~kpc of the center of the 
host halo.  In the bottom row, we compute the best-fit plane 
with respect to all luminous subhalos within 
$300 kpc$ of the host halo center.  
}
\label{fig:satplanes}
\end{figure*}

We determine the principal axes of the three simulated 
host halos and the corresponding principal axis 
ratios $q \equiv b/a$ and $s \equiv c/a$ ($a>b>c$) 
in the following way.  
We construct a modified inertia tensor given 
by \citep[e.g.,][]{dubinski_carlberg91}
\beq
\label{eq:itensor}
I_{ij} = \sum_{\nu} m_{\nu} x_{i}^{\nu}x_{j}^{\nu}/\zeta_{\nu}^{2},
\eeq
where $m_{\nu}$ is the mass of the $\nu$th particle, 
$x_{i}^{\nu}$ is the $i$ coordinate with respect to a 
reference frame centered on the density peak of the halo, 
$\zeta_{\nu}^{2} \equiv (y_1^{\nu})^2 + (y_2^{\nu}/s)^2 + (y_3^{\nu}/q)^2$, 
and $y_{i}^{\nu}$ are the particle coordinates with respect to 
the halo principal axes.  We use an iterative algorithm 
to determine the principal axes.  
We begin with the assumption of a spherical 
configuration ($a=b=c$), construct the inertia tensor according to 
Equation~(\ref{eq:itensor}), and diagonalize the tensor to determine 
the principal axes (eigenvectors) and the axis ratios (ratios of 
eigenvalues).  We then repeat this process, 
using the results of the previous iteration to define the 
principal axes, until the results converge to a fractional 
difference of $10^{-3}$.  The process generally takes 
fewer than $10$ iterations to converge.  
The factor of $\zeta_{\nu}^{-2}$ in Eq.~(\ref{eq:itensor}) 
serves to mitigate the influence of massive 
substructures at large distances, which can be a 
significant source of noise in the measurement 
\citep{dubinski_carlberg91}.  

In Figure~\ref{fig:shapes}, we show shape profiles 
at $z=0$ and $z=1$ for each host halo 
constructed in this way.  The Figure shows 
the axis ratios as a function of the length of the major 
axis of the ellipsoid $\zeta$, and we construct profiles 
by considering all particles within this ellipsoid.  
In agreement with previous studies 
\citep[e.g.,][]{jing_suto02,bullock02}, all three MW-sized 
host halos are triaxial with $b/a \approx 0.6-0.7$ and tend to 
be more prolate than oblate with $c/b \approx 0.8-0.95$.  
Additionally, it is evident that halo shapes evolve 
as halos tend to be less spherical at high redshift, 
reflecting their younger dynamical age.  

Currently, it is unclear whether such axis ratios are consistent with
observational constraints on the shape of the MW halo.  Recent studies
suggest that the coherence of the Sagittarius tidal debris constrains
the minor-to-major ratio to $c/a \gsim 0.8$ \citep{ibata_etal01,
majewski_etal03, johnston_etal04}.  However, \citet{helmi04a,helmi04b}
argues that the Sagittarius tidal stream is consistent with a
minor-to-major axis ratio as small as $c/a \simeq 0.6$ \citep[see
also][]{martinez_delgado_etal04} because the stream is dynamically
young. Moreover, the shape constraint could be relaxed in a prolate
halo where orbits along the long-axis are less susceptible to strong
precession \citep[see, however,][]{johnston_etal04}.  We note that the
effect of baryon cooling during galaxy formation should make halos
more spherical compared to halos in the dissipationless simulations
analyzed here.  The change in axis ratios is $\Delta (c/a)\sim
0.1-0.3$ in the inner regions of the halo, and it is not uncommon to
see significant variation in axis ratios with $\zeta$
\citep{dubinski94,kazantzidis_etal04}.

In our analysis below, we refer to subhalo positions in a coordinate
system defined by the principal axes of the host halo inertia tensor,
calculated using DM particles within the ellipsoid with 
$\zeta = 0.3 \Rvir$ 
(for the MW-sized halos, this corresponds to $\zeta \approx 85
\hkpc \approx 120 \kpc$).  We choose particles 
with $\zeta < 0.3 \Rvir$ in order to mitigate the 
influence of large substructures at large halo-centric distances 
on the reference frame definition; however, we 
find that the ellipsoids defined at different values of 
$\zeta$ are closely aligned at all radii within any single 
host, in agreement with \citet{jing_suto02}.  
We define the zenith angle, $0 \le \theta \le \pi/2$, 
as the angle from the major axis of the halo.  We also make 
analogous computations in two-dimensional projection, 
in which case we define a polar coordinate system with 
angle $\phi$, defined as the angle away from 
the major axis of the two-dimensional projected 
DM distribution.

\begin{figure*}[t]
\epsscale{2.0}
\plotone{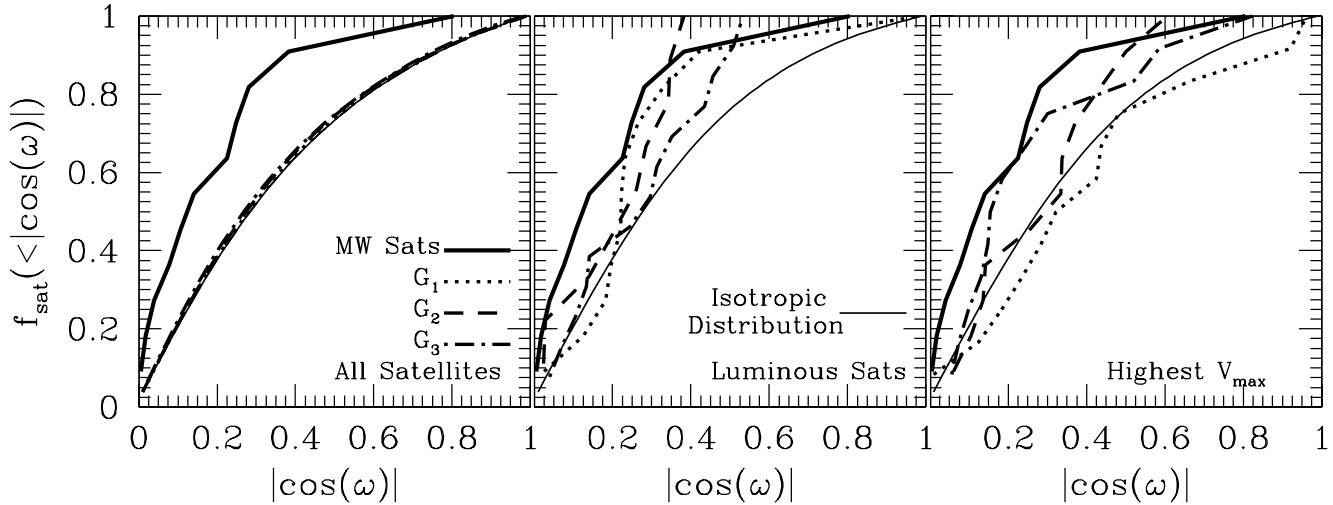}
\caption{
The cumulative fraction of halos with an angular 
position $< \acosw$ with respect to the 
normal to the best-fit plane as a function of $\acosw$.  
The {\em left} panel shows the distribution 
for all subhalos with $\Vmax \ge 12 \kms$, 
the {\em center} panel shows the distribution of 
``luminous'' satellites (see \S~\ref{sub:lumsats}), 
and the {\em right} panel shows the distribution for 
the $11$ subhalos with the largest $\vmax$ in each host.  
In each panel, the {\it dotted} lines represent 
halo $\gone$, the {\it dashed} lines represent halo 
$\gtwo$, and the {\it dot-dashed} lines represent 
halo $\gthree$.  The {\em thin, solid} lines 
in each panel represent the expected CDF for an 
isotropic satellite distribution.
}
\label{fig:Nltcw}
\end{figure*}

\subsection{Satellite Planes}
\label{sub:satplanes}

Instead of defining satellite positions in the coordinate system set
by the mass distribution of their host halo, which is difficult to
determine observationally, one can construct coordinate
systems with respect to the satellites themselves.
\citet{kroupa_etal04} found the MW satellites to be in a 
nearly planar distribution, so we follow the 
method of \citet{kroupa_etal04} and find 
{\it best-fit planes} to the satellite positions in our host halos.
Specifically, we determine a best-fit plane by minimizing the 
root-mean-square ({\sl rms}) of the perpendicular distances of all 
satellites to the plane,
\beq
\label{eq:d2plane}
\drms = \sqrt{
\frac{\sum_{\mu=1}^{N} (\hat{n} \cdot \vec{x}_{\mu} - d)^2}{N}}.
\eeq
In Eq.~(\ref{eq:d2plane}), $\hat{n}$ is a unit vector normal 
to the plane, $d$ is the perpendicular distance from the origin 
(in this case, the origin is the Galaxy or the halo center) 
to the plane, $\vec{x}_{\mu}$ is the position of the 
$\mu$th satellite, and the sum is over all $N$ satellites.  
We define a polar angle $0 \le \omega \le \pi/2$, 
as the angle between a satellite position vector $\vec{x}_{\mu}$, 
and the best-fit unit vector $\hat{n}$, 
set at the point on the plane that minimizes the 
distance to the origin. 
Additionally, the value of $\drms$ itself can also be 
used as a measure of the planarity of the satellite 
distribution \citep{kang_etal05}, 
and we consider this statistic below. 

%
%
%
%
\section{Results}
\label{sec:results}

\subsection{Visual Impression}
\label{sub:visual}

Figure~\ref{fig:satplanes} gives a visual impression of the planarity
of the observed MW satellites, as well as the subhalos and the
luminous subhalos in our simulations.  Here and below, we consider all
eleven known MW satellites within $R_{\mathrm{max}} = 300 \kpc$
\citep{grebel_etal03}.  The MW satellites are shown in the rightmost
panels of Fig.~\ref{fig:satplanes}. The planarity of the distribution,
noted by \citet{kroupa_etal04} is very clear. Note, however, that
recent studies indicate that the poles of the orbits of Ursa Minor
\citep{piatek_etal05}, as well as Sculptor and Fornax (Piatek et al.,
in preparation), are not coincident with the pole of the plane of MW
satellites, nor are they coincident with each other.  This is an
indication that the high degree of planarity of the MW dwarfs may be
transitory.

In the top row of Fig.~\ref{fig:satplanes}, we 
show projections of all subhalos of each host onto the plane
perpendicular to their best-fit plane.  In these projections the
best-fit plane runs vertically along the axis $x = 0$ and we show the
projected direction of the major axes in each panel by a thin solid
line.  In the bottom row, we show similar projections for the
luminous subhalos.  The spatial distributions of the highest-$\Vmax$
subhalos are similar to those of the luminous subhalos.

The most obvious feature in all of the panels is that the best-fit
plane of subhalos is strongly correlated with the major axis of the DM
distribution of the host halo.  If we draw $10^5$ random permutations
of eleven subhalos from the full sample, the probability that the
best-fit plane of any such subsample is inclined by more than
$45\degrees$ from the host halo major axis is only $P \simeq 9\%$,
signaling anisotropy of the subhalo population as we discuss in more
detail below.  Correspondingly, the best-fit planes for the luminous
subhalos are also aligned with the major axes of their hosts. Their
degree of planarity is visually comparable to that of the MW
satellites.    In the following subsections, 
we quantify the anisotropy of satellite distribution.

\subsection{Satellite Planarity}
\label{sub:planes}

\begin{figure*}[t]
\epsscale{2.0}
\plotone{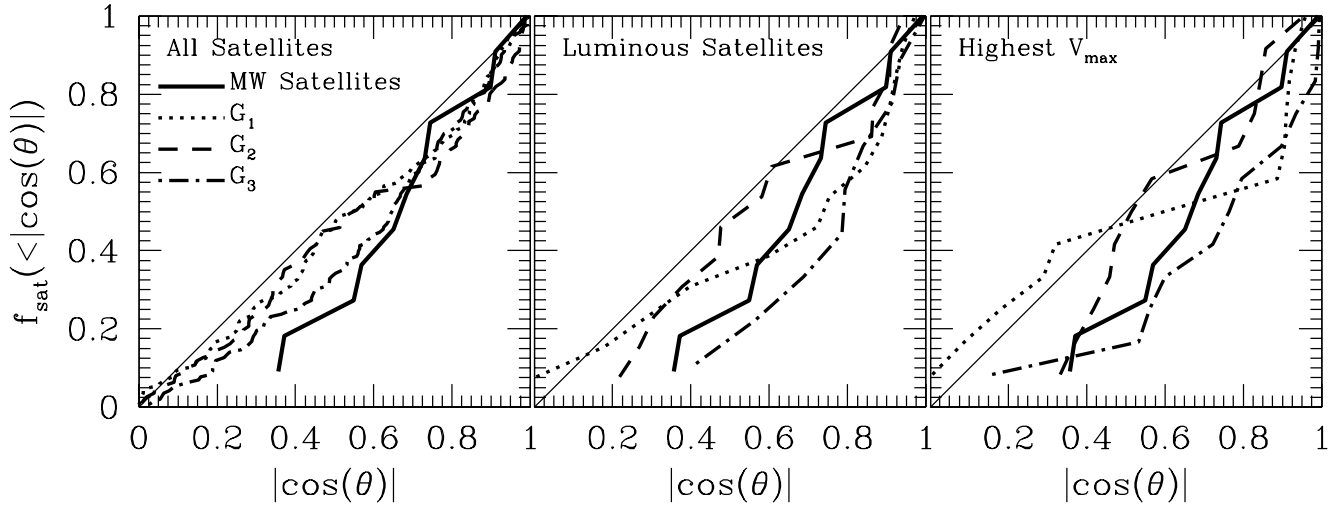}
\caption{
  The cumulative fraction of satellites with the absolute value of the
  cosine of the zenith angle $< \acost$.
  The zenith angle, $0 \le \theta \le \pi$, is defined as the angle
  from the major axis of the DM distribution of the host. 
  The {\em left} panel shows the distribution 
  for all subhalos with $\Vmax \ge 12 \kms$.  The {\em center} 
  panel shows the distribution of only those subhalos that
  are ``luminous'' according to the model of \citet[][see
  \S~\ref{sub:lumsats}]{kravtsov_etal04}.  The {\em right} panel shows
  the distribution for the eleven subhalos with the largest $\Vmax$ in
  each host.  The observed MW satellites ({\em thick solid line}) are
  placed on this plot by defining $\theta$ to be the angle from the 
  pole of the MW disk.  In each panel, the {\it dotted} 
  lines represent halo $\gone$, the {\it dashed} 
  lines represent halo $\gtwo$, and the {\it dot-dashed} 
  lines represent halo $\gthree$. }
\label{fig:Nltct}
\end{figure*}

\citet{kroupa_etal04} argued that the planarity of the observed MW
satellites is inconsistent with the CDM paradigm based on their
distribution in $\acosw$.  However, their conclusion is incorrect for
two reasons.  First, as we show below, CDM does not predict that
dwarf-sized subhalos are distributed isotropically.  Second, the null
hypothesis for the isotropic distribution for $\acosw$ used by
\citet{kroupa_etal04} is correct only in the limit of large sample
size.  This is because for a small number of objects, the distribution
of $\acosw$ is not related to a fixed reference frame, but a frame
determined by the selected objects themselves.  This modifies the
underlying cumulative distribution function (CDF) for
$\acosw$. \citet{kroupa_etal04} computed the distribution of $\omega$
for a {\em single} sample of $10^5$ subhalos, the result of which is a
nearly uniform distribution in $\acosw$ on the interval $[0,1]$. 
However, the $\acosw$ CDF for small number of objects drawn from
the isotropic distribution is different. For
example, consider the limit of randomly selecting precisely 
three objects from any underlying distribution.  
The three objects will always lie on the best-fit plane, 
such that $\acosw = 0$ for all objects,
{\it independent of their underlying distribution}.

In order to account for the small sample size, we generate the
distribution in $\acosw$ that should be expected from 
drawing eleven satellites from a uniform distribution.  
First, we assume an underlying isotropic subhalo 
distribution with a radial number density profile 
$n(r) \propto 1/(1+(r/r_{\mathrm{c}})^3)$ 
with $r_{\mathrm{c}} = 0.25 \Rvir$ 
\citep[see][who found this to be a good description of subhalo 
distributions in simulations and their semi-analytic models]{zentner_etal05}.  
Our results are not sensitive to the particular 
assumptions that we make about the 
radial distribution.  We then draw $10^5$ random samples of 
eleven satellites from this distribution, determine the 
best-fit plane for each sample, 
and compute $\cosw$ for each satellite and the 
corresponding CDF.  The resulting average CDF of 
$\acosw$ is shown as the thin, solid line in 
Figure~\ref{fig:Nltcw}.  
Notice that the $\acosw$ distribution for eleven 
satellites selected from an isotropic distribution 
is significantly different from a uniform distribution 
in $\acosw$, which would be a diagonal line.  

We can assess the probability that the observed MW satellites 
are drawn from an isotropic distribution using the CDF of 
$\acosw$ constructed as described above.  
The Kolmogorov-Smirnov (KS) probability to draw the
observed distribution of MW satellites 
within $R<300$~kpc from an isotropic distribution is 
$\Pksiso \simeq 0.15$.\footnote{If we were to follow 
\citet{kroupa_etal04} and use the uniform CDF that 
results from a very large sample size, 
we find $\Pksiso \approx 2 \times 10^{-3}$, 
in good agreement with their analysis.} 

In a similar analysis, we draw $10^5$ random samples of 
eleven subhalos from the full, simulated subhalo populations 
($\Vmax \ge 12 \kms$ and $R< 300 \kpc$) and construct
the CDFs for these samples as shown in Fig.~\ref{fig:Nltcw}.  
Comparing this distribution to the isotropic case 
gives a KS probability of $\Pks \simeq 0.49$. However, 
as we noted above, the best-fit planes in the 
subhalo samples are inclined by less than $45\degrees$ with 
respect to the major axis of the host halo in $91\%$ 
of the samples, while there is no such preferential 
alignment in the isotropic case.  
For the luminous subhalos, the KS probability that 
they are drawn from an isotropic distribution is 
$\Pks \simeq 0.39$.  Finally, 
the probability that the MW 
satellites are drawn from the distribution of all 
subhalos is $\Pks \simeq 0.16$.  
Figure~\ref{fig:satplanes} and the analysis below 
show that the distributions of both the MW satellites 
and subhalos are actually anisotropic.  
The conclusion is that 
{\it for a small sample (eleven), the 
$\acosw$ distribution is non-discriminatory}.  
Other statistics must be used 
to draw meaningful conclusions.

The problem of the reference frame uncertainty for small 
sample sizes can be avoided if the frame is fixed and 
does not depend on the objects.  In particular, 
we can choose to define the angles 
with respect to the halo major axes, 
(i.e., the zenith angle $\theta$, defined in \S~\ref{sub:inertia}).  
We show the CDF of $\acost$ for the simulated subhalos 
in the left panel of Figure~\ref{fig:Nltct} and the 
CDF for an isotropic distribution by the thin diagonal line.  
For all three of the simulated MW-sized hosts, 
the subhalos are clearly distributed anisotropically 
with a preferential alignment along the host halo major axis.  
The KS probability of drawing this subhalo population 
from an isotropic distribution is $\Pksiso \simeq 1.5 \times 10^{-4}$.
This is in contrast to the result of \citet{willman_etal04}, who find
an isotropic distribution of subhalos in a MW-sized halo simulated by
\citet{reed_etal03}.  We find the differential fraction of satellites
per unit $x \equiv \acost$ averaged over all three host halos to be
sharply peaked and well described by
\beq
\label{eq:dfdx}
\frac{\dd \fsat}{\dd x} \simeq 0.80 + 1.15 x^{4.75}.
\eeq

The thick, solid lines in each of the panels of Figure~\ref{fig:Nltct}
show a similar distribution in zenith angle for the eleven observed MW
satellites within $300 \kpc$.  The orientation of the MW halo is
unknown, so the MW satellites cannot be placed on this plot without
making an assumption about the orientation of the halo with
respect to the disk.  In the reference frame in which the
zenith angle $\theta$, is measured with respect to the Galactic pole 
(we discuss implications of this assumption in \S~\ref{sec:disc}),
the KS probability of drawing the MW satellites from a
distribution that is isotropic about the Galactic center is 
$\Pksiso \simeq 0.07$, while the probability of drawing the MW 
satellites from the sample of CDM subhalos is $\Pks \simeq 0.34$.  
Alternatively, if we assume that the MW satellites are 
aligned along the minor or intermediate axes, the 
probabilities for the subhalos and MW satellites 
to have the same parent distribution are 
$\Pks \simeq 0.02$ and $\Pks \simeq 0.03$,
respectively.

For the subhalos that are luminous according to the KGK04 model, the
preferential distribution at low zenith angles remains and the KS
probability of for all luminous subhalos to be drawn from 
an isotropic underlying distribution is only
$\Pksiso \simeq 0.02$.  For the eleven highest-$\Vmax$ subhalos 
in each host, the KS probability of being sampled from 
isotropic distribution is $\Pksiso \simeq 0.05$.
Note that for such small samples of objects, the halo-to-halo 
variation in the zenith angle distribution is large.

A useful measure of planarity is the {\sl rms} distance to 
the best-fit plane, $\drms$ 
\citep[][see Eq.~(\ref{eq:d2plane})]{kang_etal05}.  
In agreement with \citet{kroupa_etal04},
we find that $\drms^{\mathrm{MW}} \simeq 26.4 \kpc$ for the eleven MW
satellites within $300 \kpc$.  For eleven satellites selected randomly
from the isotropic distribution above, the mean value of the 
rms plane width in $10^5$ sample realizations is 
$\drms^{\mathrm{iso}} \simeq 72 \kpc$ 
and the probability of a distribution similar to the
MW satellites is 
$P(\drmsiso \le \drmsmw) \simeq 5 \times 10^{-3}$.
Similarly, for the subhalo realizations averaged over 
all three host halos, 
the mean width is $\drms \simeq 58 \kpc$ and 
$P(\drms \le \drmsmw) \simeq 0.02$.  
This shows that both MW satellites and DM
subhalos are distributed anisotropically, 
but the observed distribution is somewhat 
more planar.

The radial distribution of observed satellites is, however, 
more centrally concentrated than that of the subhalos.  The 
median distance of the MW satellites from the Galactic center 
is a factor of two smaller than the median distance of subhalos from 
the centers of their hosts \citep{taylor_etal03,kravtsov_etal04}. 
The luminous subhalos have a radial distribution that is similar 
to the MW satellites and represent a biased 
subsample of the overall subhalo population.
The {\sl rms} distance to the best-fit plane is sensitive 
to the radial extent of the population and we 
expect more centrally concentrated populations 
of objects to have smaller values of $\drms$.

\begin{figure}[t]
\epsscale{0.9}
\plotone{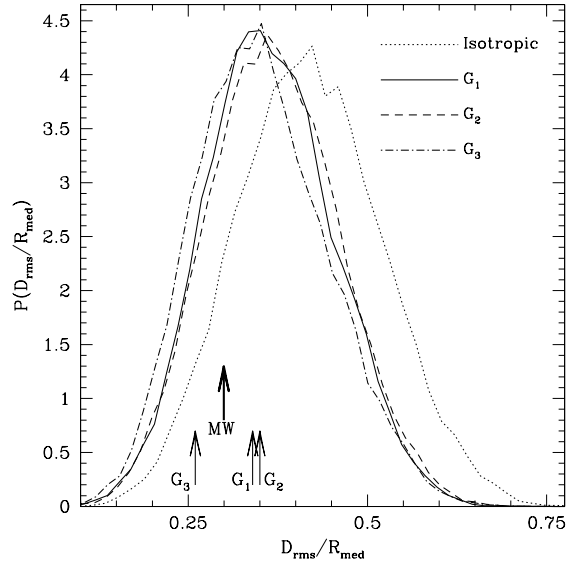}
\caption{ The distribution of the {\sl rms} dispersions of satellite
distribution around their best fit planes. The dispersions are
rescaled to the median distance of each population to the center of
their host (see text for details).  The {\em dotted} line corresponds
samples of 11 objects selected from an isotropic population
distributed radially as the simulated subhalos.  The other lines
corresponds to samples of 11 subhalos drawn from the subhalos of 
$\gone$ ({\em solid} line), $\gtwo$ ({\it dashed} line), and the
$\gthree$ ({\it dot-dashed} line).  The labeled arrows at the bottom of
the plot mark the values of $\drms/\rmed$, for the observed MW
satellites and the luminous subhalos of $\gone$, $\gtwo$, and
$\gthree$.  }
\label{fig:drms}
\end{figure}

Indeed, we find that the peak of the distribution of $\drms$ scales
linearly with the median of the radial positions of the underlying
population $\rmed$.  The rescaled {\sl rms} distance to the plane, 
$\delrms \equiv \drms/\rmed$, 
can serve as a useful measure of planarity that
normalizes out the radial extents of the different 
populations.  In Figure~\ref{fig:drms}, 
we show the probability distribution of
selecting eleven satellites with a particular value of 
$\delrms$ drawn from the isotropic distribution 
and from the samples of subhalos of hosts 
$\gone$, $\gtwo$, and $\gthree$.  For both the
isotropic distribution and for the simulated subhalos, 
$\rmed \approx 167 \kpc$.  Figure~\ref{fig:drms} 
shows that in terms of the probability distribution of 
$\delrms$, the observed value is not unusual.  
In fact, we find that the value of $\drmsmw$ is not unlikely 
($P(\delrms^{\rm iso} \le \delrms^{\mathrm{MW}}) \simeq 0.14$) 
even for a random sample of eleven objects drawn from an 
isotropic distribution, if their radii are rescaled 
to have the same $\rmed$ as the MW satellites. 
The figure also shows that the distribution of subhalos is
anisotropic, which shifts the probability distribution of 
$\delrms$, increasing the probability of selecting a 
subsample with a degree of planarity comparable to that of the 
MW satellites: 
$P(\delrms^{\rm sub} \le \delrms^{\mathrm{MW}}) \simeq 0.30$.

In the model of KGK04, the luminous subhalos of $\gone$ and $\gthree$
match the observed radial distribution of the MW satellites well.  We
thus expect that their $\drms$ will be comparable to the observed
value. Indeed, we find for the luminous subhalos that $\drms \simeq
29.4 \kpc$, $\drms \simeq 46.2 \kpc$, and $\drms \simeq 27.4 \kpc$ for
hosts $\gone$, $\gtwo$, and $\gthree$, respectively. These values are
similar to $\drmsmw$.  The values of $\delta_{\rm rms}$ for the MW
satellites and the luminous subhalos are shown by arrows in
Figure~\ref{fig:drms}.

These results indicate that {\it the main reason that the MW
satellites occupy a narrower plane than DM subhalos is their more
centrally-concentrated radial distribution}.  Given that there are
physical reasons to expect such radial bias (KGK04), we conclude that
the observed anisotropy of the MW satellites is consistent with CDM
predictions.  However, we note that this agreement requires
approximate alignment of the major axis of the halo that hosts the MW
and the pole of the MW disk.  In the next section, we discuss the
origin of the subhalo anisotropy and the implications of such
disk--halo alignment.

\section{Discussion}
\label{sec:disc}

\subsection{Origin of the Subhalo Anisotropy}
\label{sub:origin}

\begin{figure*}[t]
\centerline
{
        \epsfysize=2.75truein \epsffile{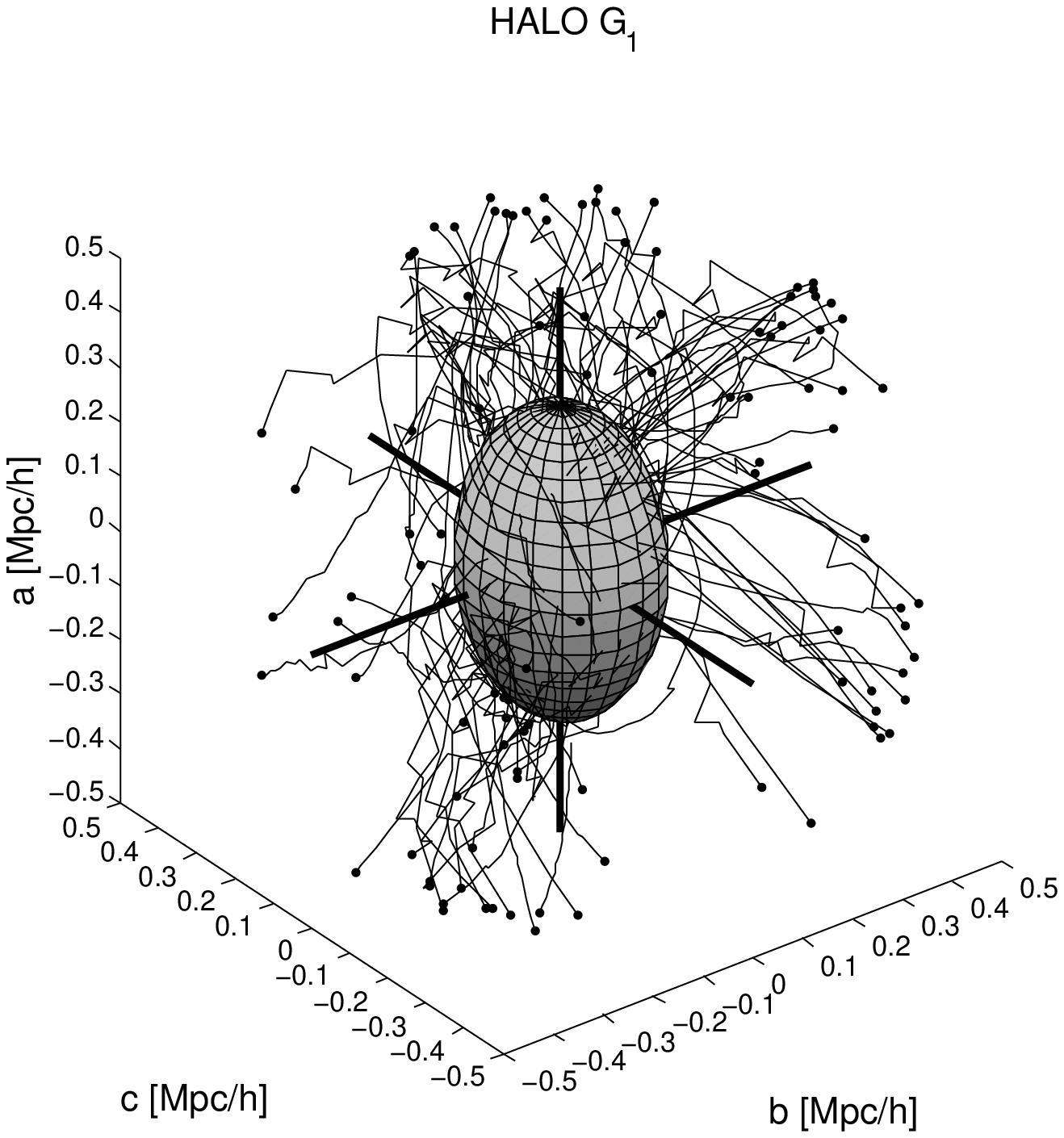}
        \epsfysize=2.75truein \epsffile{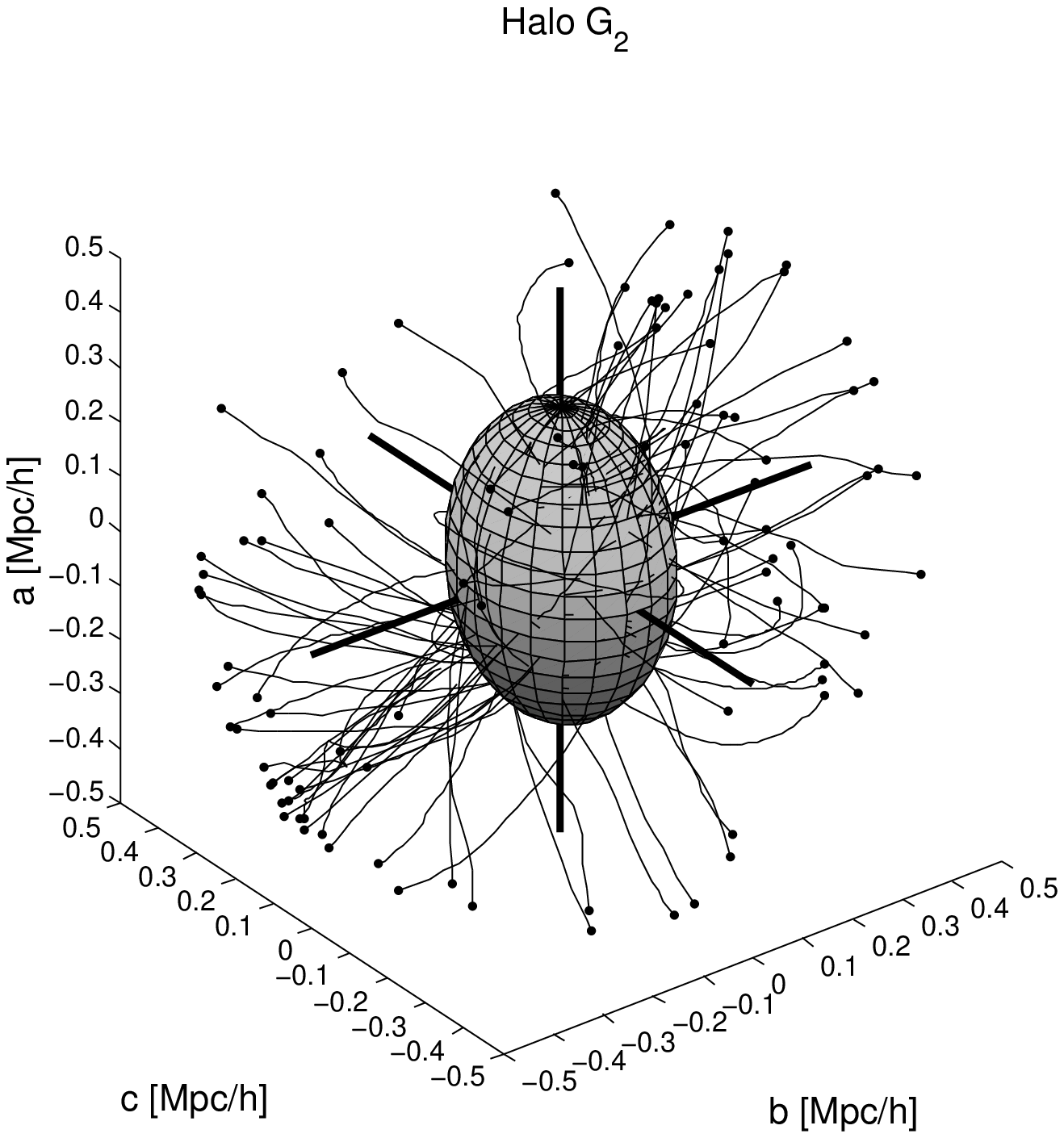}
        \epsfysize=2.75truein \epsffile{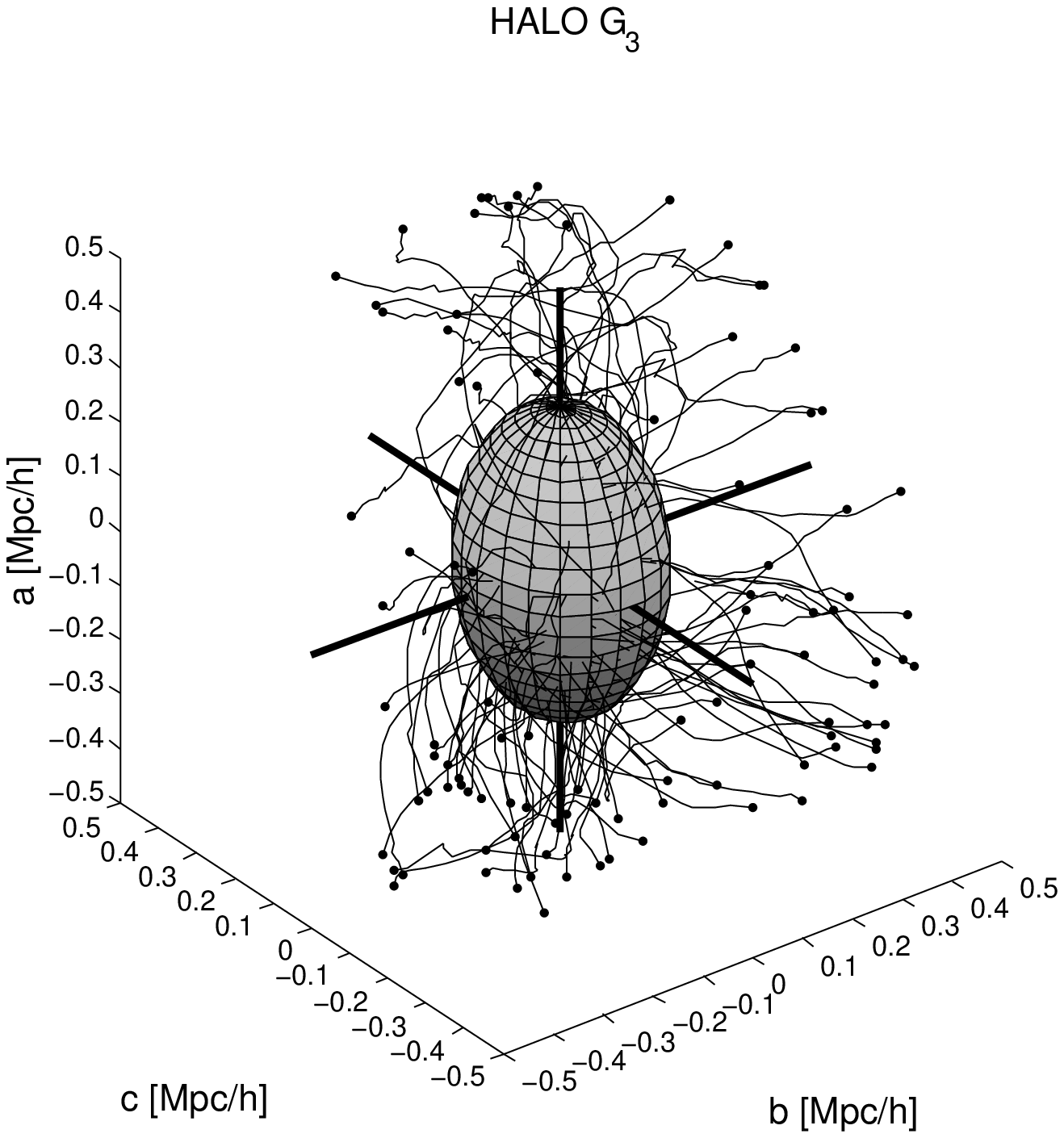}
}
\caption{
  Trajectories of 100 randomly selected subhalos accreting onto host
  halos $\gone$, $\gtwo$, and $\gthree$ 
  (panels from left to right). The trajectories are
  plotted relative to a coordinate system centered on the most bound
  particle in the parent halo and oriented along the principal axes of
  the parent halo evaluated at $z=0$. 
  The coordinates are comoving. The first
  point of each trajectory is at the time when the comoving 
  distance between subhalo and host first became 
  smaller than $600 \hkpc$.  
  In each panel, the ellipsoids have major axes of length 
  $\zeta = 290 \hkpc$, and the axis ratios are set 
  to the axis ratios of each host at 
  $\zeta = \Rvir$ and $z=0$.  
  The {\em thick, solid, mutually-orthogonal lines}
  denote the principal axes of the host halos.  We label each
  coordinate axis by the corresponding principal axis ($a > b > c$)
  with which it is aligned.  Preferential accretion along the
  direction of the long axis is evident for halos $\gone$ and $\gthree$.}
\label{fig:acc}
\end{figure*}

It is interesting to ask why satellite halos should have a 
strongly-anisotropic distribution aligned with the 
major axis of the host halo.

Several suggestions have been made to explain the anisotropy
observed by \citet{holmberg69} and \citet{zaritsky_etal97}.
\citet{quinn_goodman86} investigated the effect of enhanced dynamical
friction for orbits that are nearly co-planar with a galactic disk.
The idea is that nearly co-planar orbits would be driven toward the
disk plane \citep[e.g.,][]{binney77} and decay more rapidly due to 
additional interactions with the disk component.  In this way, 
satellites on nearly co-planar orbits would be preferentially
cannibalized by the disk and these orbits depopulated.  For
example, such cannibalized dwarf galaxies can significantly contribute
to the formation of the thick disk \citep{abadi_etal03}. However, the
conclusion of \citet{quinn_goodman86} was that this process is not
efficient enough to account for the results of
\citet{holmberg69} for satellites closer than $\sim 50 \kpc$.
\citet{penarrubia_etal02} extended this argument to include the effect
of an oblate DM halo.  They found this process to be
efficient only within $\sim 50 \kpc$ of the host due 
to considerably longer orbital decay times at larger radii 
\citep{zaritsky_white94, zaritsky_gonzalez99}.  
Note that the MW-sized halos in our
simulations are prolate, so this process should be 
even less efficient. Yet, 
the anisotropy of satellites is present.

Two possibilities are that the anisotropy reflects a direction of
preferential infall set by the environment \citep[e.g.,][]{tormen97}
and/or that there is some other dynamical process that drives 
evolution toward anisotropy after accretion onto the host halo.
\citet{knebe_etal04} investigated the first possibility for cluster
halos to address the anisotropy observed in systems such as the Virgo
cluster \citep[e.g.][]{west_blakeslee00}.  \citet{knebe_etal04}
concluded that preferential accretion along the directions of
filaments accounts for much of the bias in satellite orbits in
cluster-sized systems.  This result does not extend trivially to
MW-sized systems.  Clusters are rare, highly-biased objects that
generally form relatively recently at the ``nodes'' of filaments 
that are comparable in thickness to the size of clusters themselves
\citep[e.g.,][]{klypin_shandarin83, bond_etal96, colberg_etal05}.  It
is therefore not surprising that many merging halos get ``funneled''
into clusters along dominant filamentary directions.  Unlike clusters, 
MW-sized objects typically form earlier, are significantly less 
biased at the present epoch, and {\em do not} generally reside in
filaments of comparable dimension.  In this case, it seems that
infalling substructure may be less likely to accrete along a single,
dominant filamentary direction over the formation history of the halo.
In addition, clusters have generally assembled their masses and
accreted their satellites much more recently than galaxy-sized systems.
As a result, many cluster subhalos are dynamically young and have
undergone only a small number of orbits ($\lsim 1-2$) in the potential
of the main halo, while the subhalos in our MW size halos have
typically undergone several orbits within the main halo 
(see, e.g., KGK04).

\begin{figure*}[t]
\epsscale{1.65}
\plotone{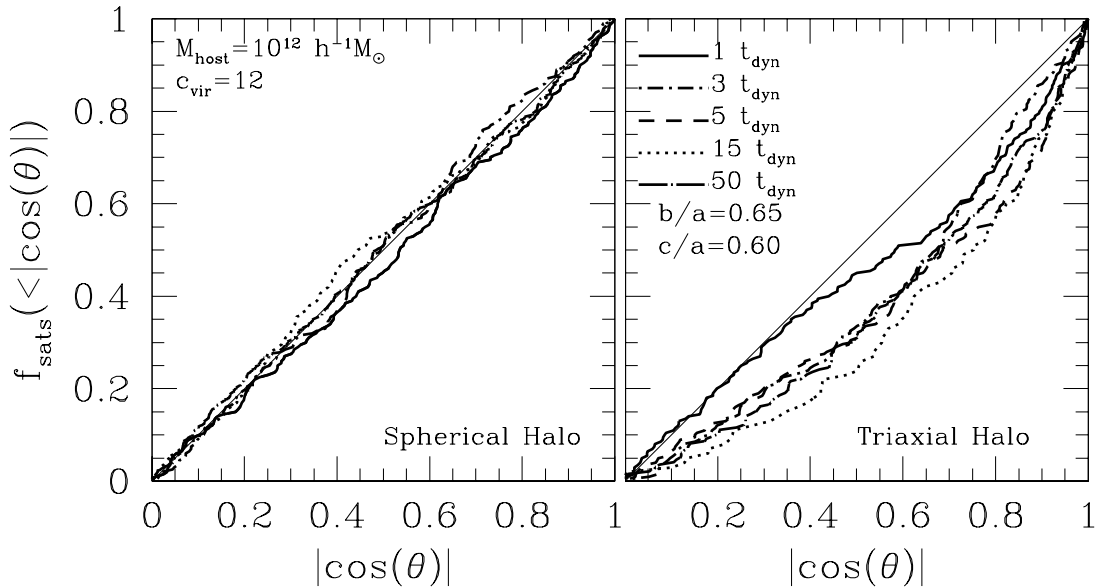}
\caption{
  Generation of an anisotropic distribution of satellites in triaxial
  potentials.  The distribution of zenith angle cosines for 200 
  isotropically accreted test particles at different epochs 
  ({\em lines of various type}; 
  the time is in units of the dynamical time of the host at
  the virial radius).  The {\it left panel} represents control case
  of the spherical halo potential, while in the {\it right panel} the
  host has ellipsoidal isodensity contours with axis ratios of
  $b/a=0.65$ and $c/a=0.60$, similar to those of 
  halos $\gone$, $\gtwo$, and $\gthree$.  The
  fiducial host halo has a virial mass of $\Mvir=10^{12}\,\hMsun$ and
  the NFW mass distribution with concentration parameter of
  $\cvir=12$.  The {\it solid} line indicates an integration for
  $1\,\tdyn$, the {\it dot-short-dashed} line indicates integration for
  $3\,\tdyn$, the {\it dashed} line for $5\tdyn$, the {\it dotted} line
  for $15\,\tdyn$, and the {\it dot-long-dashed} line for $50\,\tdyn$.
  The {\it thin solid} line corresponds to the isotropic distribution.
  The figure shows that in the spherical potential tracer distribution
  remains isotropic at all times.  In this case of prolate, triaxial
  potential, the orbits evolve toward an anisotropic distribution
  similar to that of the observed satellites and simulated subhalos.
}
\label{fig:gen}
\end{figure*}

In Figure~\ref{fig:acc}, we show trajectories for $100$
randomly-selected subhalos of each host halo as they are accreted.
The trajectories are constructed from the $96$ saved simulation
timesteps by examining the $25\%$ most bound particles in each halo.
Halos at adjacent timesteps that share the highest common fraction of
these particles are identified with each other, as described in \S~4
of KGK04.  In two of the three cases (halos $\gone$ and $\gthree$),
there is clearly preferred satellite accretion along the direction of
the major axis of the host halo.  In the case of halo $\gtwo$, the
accretion is less anisotropic, but there appears to be a small
preference for accretion toward the octant $a<0$, $b<0$, $c>0$.  We
have examined the accretion histories of the host halos more closely
and halo $\gtwo$ appears to feed off of a filament that runs roughly
in this direction during its early evolution ($z \gsim 1$).  It
subsequently accretes along a filament that is more closely aligned
with its major axis direction at $z=0$.  
Note that the halos $\gone$ and $\gthree$ have both 
the highest degree of satellite anisotropy 
(see \S~\ref{sec:results}) and the most pronounced 
preferred accretion directions.

These results indicate that a preferential direction of satellite
accretion is an important factor in determining the distribution of
satellites in MW-sized halos, just as it is for cluster-sized hosts.
Although filaments at the present time may be thick, they were
significantly thinner in the past, when many of the surviving subhalos
were accreted.  Moreover, the matter distribution in filaments 
is concentrated toward the axis of the 
filament \citep[e.g.,][]{colberg_etal05}. Finally, a 
fraction of subhalos may be accreted as members of groups, 
which are biased spatially and often located near the 
centers of filaments.  That the preferential
accretion direction is correlated with the major axis 
of the host halo is not surprising, because the major 
axis is typically determined by the 
direction of the most recent major merger.

In our DM-only simulations, effects like the cannibalization
of satellites on co-planar orbits by material associated with the disk
are absent, yet the substructure anisotropy is present, even for
subhalos that have orbited within the host potential for many
dynamical times.  This is due to the fact that elongated potentials,
similar to the potentials induced by our triaxial host halos, support
orbits that make long excursions along the major axis of the potential
\citep[see][]{statler87}.  To illustrate this, we perform a simple,
idealized experiment.  We integrate the orbits of $200$ test
particles in the static potential of a triaxial generalization 
of the density profile of 
\citet[][NFW hereafter]{navarro_etal97},

\beq
\label{eq:triaxialnfw}
\rho(\zeta) \propto \Bigg( \frac{\zeta}{\rs} \Bigg)^{-1}
\Bigg(1 +  \frac{\zeta}{\rs} \Bigg)^{-2}, 
\eeq
with ellipsoidal contours of constant density.  
$\zeta$ is defined as in \S~\ref{sub:inertia}.
For simplicity, we choose constant axis ratios 
$b/a=0.65$ and $c/a=0.6$, similar to our simulated 
host halos (Fig.~\ref{fig:shapes}), and a 
halo concentration $\cvir = \Rvir/\rs = 12$, 
which is typical of MW-sized halos \citep{bullock_etal01}.  
We choose the initial velocities for each orbit according 
to the distribution of initial conditions for 
subhalos presented in \citet{zentner_etal05}.  
In order to demonstrate the influence of the triaxial 
potential, we assumed {\em spherically symmetric} 
infall, rather than the anisotropic infall depicted in 
Figure~\ref{fig:acc}.

The results of this exercise are shown in
Figure~\ref{fig:gen}, where we plot the distribution 
of subhalo zenith angle for the subhalos in 
the triaxial model and a spherical model that 
serves as a control.  We integrate the orbits 
over an interval of $50 \tdyn$, 
where $\tdyn$ is the dynamical time 
of the model at $\Rvir$.  The net effect is clear and 
not surprising, but is often neglected.  
The prevalence of orbits that make long excursions 
along the major axis of the potential induces and 
maintains an anisotropic distribution of test particles.
The test particles assume a distribution consistent 
with the triaxiality of the potential and, 
in fact, in a live, triaxial density distribution, 
it is the presence of these orbits that 
maintain the triaxial shape of the system 
\citep{gerhard_binney85, udry_martinet94, 
barnes_hernquist96, merritt_quinlan98, 
valluri_merrit98}.

We neglected many effects in this experiment.  
These include the influence of the subhalos on 
the host potential, and the growth and 
evolution of the host halo potential 
that occurs in cosmological simulations.  
As we show in Figure~\ref{fig:shapes}, the 
host halo shapes evolve from 
$c/a \sim 0.4$ at $z \sim 1$ to 
$c/a \sim 0.6$ at $z \sim 0$.  
Moreover, subhalos on elongated orbits will be more
vulnerable to tidal disruption because they generally pass 
closer to the center of the potential where the tides are 
strongest \citep[e.g.,][]{zentner_bullock03,zentner_etal05}.  
The net effect should be that such orbits will be gradually 
depopulated.  Nevertheless, our experiment 
illustrates that regardless of anisotropic infall, 
the satellite distribution can develop anisotropy due 
to the triaxiality of the host potential, 
even in the absence of any enhanced destruction
due to the central disk.

\subsection{Anisotropy of the globular cluster distribution}
\label{sub:gc}

As we pointed out above, our results indicate that the observed
distribution of the MW dwarf satellite galaxies is generally 
consistent with CDM predictions if the major axis of the 
MW DM halo is approximately aligned with the 
Galactic pole.  If this is the case, based on our
discussion of orbital structure in the previous section, 
we should expect similar anisotropies to exist in
other populations that may serve as test particles in the 
MW halo potential.  

Consider the distribution of the MW globular clusters (GCs).
\citet{frenk_white82} studied a subsample of the MW GCs and
concluded that both metal-rich and metal-poor globular cluster systems
are slightly flattened.  Subsequently, \citet{zinn85} used a larger GC
sample and concluded that metal-rich GCs are in a disk-like
configuration \citep[see also][]{armandroff89}, while the distribution
of the metal-poor clusters is nearly spherical.  More recently, 
\citet{hartwick00} analyzed the distribution of $15$ metal-poor, 
distant globular clusters ($R \gsim 25 \kpc$ and $ \feh < -1$) 
and found that they form a flattened system with a minor axis 
highly-inclined relative to the MW disk rotation axis.

\begin{figure}[t]
\epsscale{0.9}
\plotone{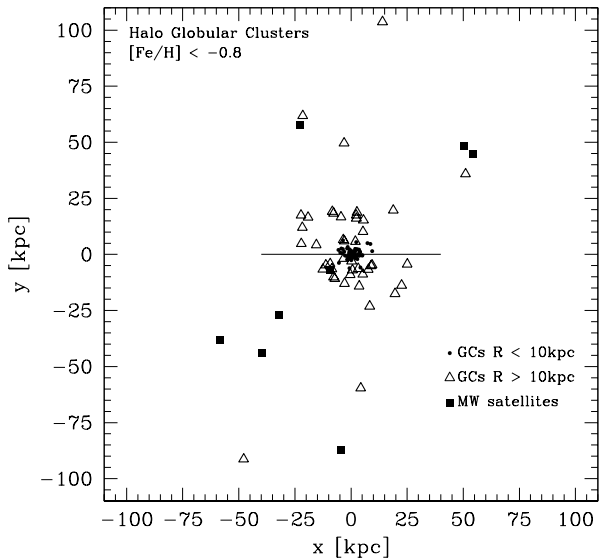}
\caption{
  The distribution of metal-poor, MW globular clusters ($\feh < -0.8$)
  and dwarf satellite galaxies. All positions are projected 
  onto a plane perpendicular to the best-fit plane of the $44$ most 
  distant globulars ($R>10\kpc$).  The {\em points} correspond to 
  the GCs with $R<10\kpc$.  The {\em open triangles} correspond to 
  GCs with $R>10\kpc$.  For comparison, we show the 
  positions of the MW satellite galaxies as the {\em filled squares}.  
  The {\em thin, solid, horizontal} line represents the orientation 
  of the MW disk.  Note that only $8$ of the MW satellites fall within 
  the limits of this plot and the projection is oriented at an angle 
  of nearly $36^{\circ}$ with respect to their best-fit plane. 
}
\label{fig:GC}
\end{figure}

Following \citet{zinn85}, we divide the GC sample of
\citet{harris96}\footnote{The globular cluster catalog is available at 
URL {\tt http://physwww.physics.mcmaster.ca/$^{\sim}$harris/mwgc.dat}.}
into halo and disk clusters using a metallicity threshold.  
We consider the distribution of metal-poor 
($[ \mathrm{Fe/H} ] \le -0.8$), {\em halo GCs}.
The distribution of the metal-poor GCs is shown in 
Figure~\ref{fig:GC} along with the distribution of the 
innermost MW dwarf satellites.  Visually, the distribution 
of these two populations of halo objects appear
to have similar anisotropy. 

Just as for the dwarf satellites, we compare the distribution of GCs
to an isotropic distribution using the distribution of $\acost$,
where $\theta$ is defined relative to the Galactic pole, 
and $\delrms^{\mathrm{GC}}$ because 
$\acosw$ has little discriminatory power.  The results
are summarized in Table~\ref{tab:gc}.  The inner globular 
clusters ($R < 10 \kpc$) 
appear to lie in a flattened distribution 
($P(\delrms^{\mathrm{iso}} < \delrms^{GC}) \simeq 0.02$) 
that is closely aligned with the disk plane.  
The normal to the best-fit plane of inner halo GCs 
is offset form the MW pole by only $\Tmw \simeq 17\degrees$.  
Similar to \citet{hartwick00}, we find that the distant
halo GCs show marginal evidence for a distribution that is 
aligned with the pole of the MW disk 
($\Tmw \simeq 80\degrees$), a configuration that is 
consistent with the distribution of 
the MW dwarf satellites.

%
%
%

An additional piece of evidence comes from the distribution 
of the M31 satellites.  In this case, distances to the 
M31 subgroup have considerably larger errors than the 
MW dwarfs, so these objects carry less statistical weight.
Nevertheless, \citet{hartwick00} studied the satellites of M31 
in detail and found that, like the MW satellites, they appear to 
be arranged in a prolate structure that has its best-fit 
major axis misaligned with the pole of the M31 disk by 
only $\sim 20\degrees$.  The distributions of 
M31 and MW satellites appear to show a similar 
anisotropy with preferential alignment 
along the galactic poles.

%
%
\begin{table}[t]
\label{tab:gc}
\begin{center}
{\sc Anisotropy of the metal-poor Milky Way Globular Cluster Distribution}\\[2mm]
\begin{tabular}{lcccccc}
\tableline\tableline\\
{$R$~$[\kpc]$} & 
{$\Tmw$} &
{$\Tsats$} &
{$\Pkst$} & 
{$N_{\mathrm{GC}}$} 
\\[2mm]
\tableline\\
$0-10$ & $17\degrees$ & $79\degrees$ & $0.12$ & $68$ \\
$> 10$ & $80\degrees$ & $36\degrees$ & $0.03$ & $44$ \\
\\
\tableline
\end{tabular}
\end{center}
{\small Column description: (1) the range of galacto-centric distances 
in GC subsample; (2) $\Tmw$, the angle between 
the MW pole and the unit vector normal to the best-fit 
plane of GCs in the sample ($\Tmw = 0\,\deg$ corresponds to the 
Galactic disk plane, while $\Tmw = 90\,\deg$ corresponds to a plane 
parallel to the MW pole); (3) $\Theta_{\rm SATS}$, 
the angle between the best-fit plane of GCs and the 
best-fit plane of MW dwarf satellite galaxies; 
(4) $\Pkst$, the KS probability for the GC sample to 
be drawn from an isotropic distribution using the 
cumulative distribution of $\acost$, where $\theta$ is the angle
with between position vector of the satellite and Galactic pole; 
(5) the number of halo GCs in each subsample.
}
\end{table}
%

\subsection{Implications}
\label{sub:implications}

The angular distribution of satellites and its interpretation 
in the context of the anisotropy of the CDM subhalos have 
important implications for our understanding of 
galaxy formation.  Collapsing halos acquire angular momentum 
via interactions with quadrupolar mass density fluctuations
\citep{peebles69,doroshkevich70,efstathiou_jones79,barnes_efstathiou87}.
In the simplest scenario of disk galaxy formation, 
baryons in halos begin by sharing the angular momentum 
distribution of the DM, on average, 
conserving it as they cool and condense
\citep[e.g.,][]{fall_efstathiou80}.  This leads to a picture 
where the poles of disk galaxies are collinear with the 
net angular momentum vectors of their host halos, 
which are generally aligned with halo 
{\em minor} axes \citep[e.g.,][]{warren_etal92,porciani_etal02,faltenbacher_etal05}.

\citet{vdb_etal02} studied the alignment of the angular momenta 
of DM halos and their baryons in adiabatic simulations, 
and found that they are generally well aligned, 
with a distribution for the angular misalignment that is 
sharply peaked between $\sim 10\degrees - 20\degrees$, 
but with an extended tail to larger angles such that the 
median is $\sim 30\degrees$.  \citet{chen_etal03} found similar 
results in adiabatic simulations and models with 
radiative cooling.  Both \citet{vdb_etal02} and \citet{chen_etal03} 
measured increasing misalignment with decreasing halo mass and 
speculated that the extended tail could be partially due to noise 
in the measurement of angular momenta in small objects.  
We note that \citet{kazantzidis_etal04} found the gaseous 
disk of a young ($z \gsim 4$) galaxy progenitor in their 
simulation to be aligned nearly {\em perpendicular} to the 
major axis of the DM halo, which is the alignment that 
our results suggest.

In this paper, we have shown that the predicted 
spatial distribution of CDM subhalos 
is consistent with the distribution of the 
MW satellites, {\em if} the pole of the MW disk is 
nearly aligned with the {\em major} axis of its 
outer DM halo. This requires a more complicated 
disk formation scenario where the halo and baryons mutually 
adjust as they evolve toward a stable configuration.  
It would be interesting to explore whether such an 
alignment is supported by observational evidence 
from other galaxies and numerical models of 
galaxy formation.  Along these lines, 
we show the anisotropic, two-dimensional 
projected subhalo distributions in the Appendix.

As we noted in the introduction, current observational results are
contradictory.  \citet{holmberg69}, \citet{zaritsky_etal97}, 
and \citet{sales_lambas04} report that 
satellites of other galaxies exhibit statistical
anisotropy similar to that of the MW dwarfs.  Specifically, they find
satellites to be preferentially located near the minor axes of the
projected {\em light} distributions of host galaxies.  However, the
study of \citet{brainerd04} shows evidence of the opposite correlation
of satellite position with the {\em major} axis of their host galaxy.

One potential avenue for checking the disk--halo 
alignment is weak lensing. \citet{hoekstra_etal04} 
recently presented the first weak lensing measurement of halo 
ellipticity obtained under the {\em assumption} that the halo 
mass and galaxy light distributions are aligned. This is likely 
a sound assumption for early-type galaxies, which may indeed
dominate the lensing signal in a sample of mixed morphological 
types.  However, if galactic disks are preferentially aligned 
orthogonal to the major axes of their halos, the alignment 
for late-type systems would be opposite to that assumed by 
\citet{hoekstra_etal04}. Observational tests should be 
possible with the large dataset of the Sloan Digital Sky Survey.

On the theoretical side, \citet*{navarro_etal04} recently argued that
the angular momenta of galactic disks are aligned perpendicular to the
minor axis and parallel to the intermediate axis of the inertia tensor
of the surrounding matter distribution, both in their simulations of
galaxy formation and in observations of nearby galaxies in the Local
Supercluster.  Unfortunately, it is not clear whether the 
major axes of galactic halos are always aligned with the 
intermediate axis of the surrounding large-scale structure. 
\citet{navarro_etal04} also reported that 
the disks of nearby spirals tend to be oriented perpendicular 
to the Supergalactic plane.  We generally expect the major axes 
of halos to be oriented along the filamentary structure of the 
Local Supercluster, so such a tendency is consistent 
with a scenario where disks orient themselves perpendicular 
to halo major axes.  

The shape of the halo and the orbits of satellites within prolate and
triaxial hosts may have important consequences for disk heating and
the build up of the stellar halo.  For example, given that the 
orbital energies of most surviving subhalos are relatively large, 
one may expect that the number of satellite passages close to the 
galactic center within the last $\sim 10\Gyr$ should be small
\citep[e.g.,][]{font_etal01}.  However, a satellite on a box-like 
orbit can pass arbitrarily close to the disk.  
This may enhance both the heating of the stellar disk 
\citep[e.g.,][]{toth_ostriker92, huang_carlberg97,
  ibata_razoumov98, moore_etal99, velazquez_white99} and the 
efficiency of tidal disruption of satellites 
and the formation of the stellar halo
\citep[e.g.,][]{helmi_white99,bullock_etal01}.  Objects that 
experience close passages will be preferentially 
disrupted during halo evolution, such that 
it may appear that the present-day halo is 
devoid of satellites capable of doing damage.
These considerations imply that the triaxiality of 
host potentials has to be taken into account in 
semi-analytic and numerical models of subhalo evolution 
\citep[e.g.,][]{bullock_etal00,taylor_babul01,taylor_babul04,
benson_etal04,zentner_bullock03,zentner_etal05} in order for 
the treatment to be accurate.

Lastly, the alignment of satellite orbital planes along the halo major
axis can have implications for the evolution of tidal tails.
\citet{helmi04a,helmi04b} argues that the data on the Sagittarius
tidal debris \citep{majewski_etal04, law_etal04} provide a strong
indication that the MW halo is {\em prolate} with its major axis aligned
with the Galactic pole and that the mean axis ratio within the orbit
of Sagittarius is $0.65 \lsim c/a \lsim 0.8$.  Again, this is the
alignment necessary to explain the positions of the MW dwarfs.
Note, however, that \citet{johnston_etal04} argue that the
precession of the orbital plane of Sagittarius implied by the data
require an {\em oblate} halo that is, at most, slightly flattened 
$c/a \gsim 0.85$.

\subsection{Caveats}
\label{sub:caveat}

One of the caveats to our results (and, indeed, to the results of the 
related studies of \citet{knebe_etal04} and \citet{kang_etal05}) is 
that the simulations that we present follow the dissipationless 
evolution of DM only.  The net effect of dissipation and 
the condensation of baryons on the anisotropy of satellites is 
unclear.  There are several aspects to consider.

Cosmological gasdynamics simulations of galaxy and cluster formation
show that radiative gas cooling results in DM halos that are
significantly more spherical in their inner regions than halos in
dissipationless simulations \citep{kazantzidis_etal04}.  At 
$\zeta = 0.1 \Rvir$, the average increase in the 
minor-to-major axis ratio is $\Delta(c/a) \sim 0.3$, 
but this shift is a declining function of $\zeta$, 
such that at $\zeta \gsim 0.5$ it is $\Delta(c/a) \lsim 0.1$.  
This does not affect the accretion of satellites along
preferred directions (Fig.~\ref{fig:acc}), but will make the halo
potential in the inner regions more spherical, reducing 
any alignment of orbits with the halo major axis.  However,
this effect is small at large galacto-centric radii and satellites
with large apocenters may still move under the influence of an
effectively prolate potential.

The presence of a disk could cause the DM halo in its vicinity to
adopt a locally-oblate shape aligned with the disk plane (as opposed
to the generally prolate shapes seen in cosmological $N$-body
simulations), again enhancing destruction of satellites on 
prograde, co-planar orbits due to anisotropic dynamical friction.  
If the disk is oriented perpendicular to the halo major axis this would 
{\em enhance} satellite anisotropy compared to our results.  
Finally, it is possible that the observed, strong anisotropy of 
the MW satellites is partially due to obscuration by the disk.  The
incompleteness of the current satellite sample is uncertain, though 
\citet{willman_etal04} argue that a significant fraction 
of MW satellites at large distances may still be undetected.

\section{Summary and Conclusions}
\label{sec:conc}

We study the spatial distribution of dwarf subhalos in 
MW-sized DM halos using dissipationless cosmological
simulations of the concordance flat $\Lambda$CDM cosmology.  
Specifically, we compare the simulated 
subhalo populations with the observed
distribution of the known MW satellite galaxies.  We also test
whether the predictions of CDM simulations are consistent with
observations using two possible scenarios for mapping luminous
satellite galaxies onto subhalos in MW-sized DM halos.  
In the first scenario, the luminous dwarf satellites are 
identified using the semi-analytic model for dwarf galaxy formation 
proposed by \citet{kravtsov_etal04}.  
In the second, the luminous satellites reside in
those few subhalos with the largest maximum circular 
velocities $\Vmax$ \citep{stoehr_etal02,stoehr_etal03}.  
Our main results can be summarized as follows.

\begin{itemize}
  
\item[1.] The distribution of subhalos in host DM halos is
  {\em not} isotropic.  Subhalos are preferentially aligned with the
  major axis of the triaxial host halo mass distribution.  The KS
  probability of choosing the subhalo populations in our simulated 
  MW-sized host halos from an isotropic distribution is 
  $\Pksiso \simeq 1.5 \times 10^{-4}$.
  
\item[2.] The method used by \citet{kroupa_etal04} does {\em not} 
  demonstrate that the MW satellites are inconsistent with either 
  an isotropic underlying distribution or a distribution similar 
  to that of subhalos in CDM simulations.  We argue that 
  \citet{kroupa_etal04} adopted an incorrect null hypothesis.  
  
\item[3.] In terms of angular distribution, the probability for
the MW satellites to be drawn from the anisotropic distribution of
simulated subhalos is $\Pks \simeq 34\%$ under the assumption that 
the MW pole is aligned with the major axis of its host halo.  
Alternatively, if the MW pole is aligned with the minor axis of the 
host halo, the KS probability is $\Pks \simeq 2\%$.

\item[4.] The apparent planarity of the MW satellite distribution can
be explained by the anisotropy of the subhalo distribution and the
relative radial bias of luminous dwarf satellites relative to
subhalos.  Specifically, the subhalo subsamples that correspond to the
luminous subhalos in the model of KGK04 and the eleven highest-$\Vmax$
subhalos exhibit a degree of planarity that is similar to the observed
MW satellites.  Note again that in our simulations such planar
distributions are likely to be nearly aligned the major axis of the
host halo. This, in turn, implies that near alignment of disk pole and
halo major axis is required to explain the observed satellite distribution.

\item[5.] In agreement with \citet{hartwick00}, we find that distant
  ($R \gsim 10$~kpc), metal-poor ($[{\rm Fe/H}]<-0.8$), MW globular 
  clusters exhibit anisotropy similar to that of the dwarf satellites.  

\item[6.] The observed anisotropy of the MW satellites compared with 
   the CDM predictions for subhalo orientations, 
   along with evidence for the Holmberg effect in other galaxies 
   \citep{holmberg69, zaritsky_etal97, sales_lambas04} including the 
   dwarf satellites of M31 \citep{hartwick00}, the distribution 
   of MW halo globular clusters, and the indirect arguments of 
   \citet{helmi04b}, provide evidence for a consistent picture 
   in which the outer DM halos surrounding spiral galaxies 
   should be nearly perpendicular to the disk planes.  This 
   has interesting implications for the understanding of 
   disk galaxy formation and the orbital evolution of 
   satellite galaxies.

\end{itemize}

We discuss the origin of the anisotropy of the subhalo distribution in
simulations and show that, similar to galaxy clusters, Galaxy-sized 
halos accrete substructure along a preferential direction that is 
strongly correlated with the major axis of the host halo 
(Fig.~\ref{fig:acc}).  We also stress that orbital evolution in a
triaxial potential results in an anisotropic spatial distribution of 
tracer objects, even if their accretion is isotropic. 

The fact that consistency of observations with the $\Lambda$CDM
prediction appears to require near alignment of disk angular momenta
with the major axis of host halos is surprising.  The angular momenta
of halos are typically aligned with their {\it minor} axes
\citep[e.g.,][]{warren_etal92,porciani_etal02,faltenbacher_etal05}.
This could indicate that disk formation is accompanied by the
evolution of the angular momentum of the baryonic material as it cools
and condenses in the center.  This would not be surprising given that
condensation of baryons has a significant impact on the density and
shape of the surrounding DM halo
\citep{gnedin_etal04,kazantzidis_etal04}.  However, more detailed
numerical studies of disk galaxy formation are needed to understand
the underlying processes in detail.

It would be interesting to test whether the putative disk--halo
configuration is ubiquitous.  Such tests can be attempted 
using several approaches. For example, a particular
alignment hypothesis can be tested with weak lensing measurements.
Another statistical test can be done using the projected 
distributions of satellite galaxies in large galaxy surveys.  
In the Appendix, we present the distribution of expected 
satellite angles with respect to the major axis of 
the projected host halo ellipse (a more complete study 
will be presented in a forthcoming paper, 
Zentner et al., in preparation).  
This predicted distribution can be compared 
to the observed satellite distribution around the
major axis of the {\it light} distribution of disk galaxies. Such
a comparison can test any correlation between disk orientation and dark
matter elongation.  A detection would indirectly indicate a 
preferential orientation of disk galaxies in 
DM halos, while a null result would 
have the interesting implication that the correlation between 
disk and halo orientation is, at most, weak.  
These tests should be feasible with existing 
large galaxy surveys, such as the Sloan Digital Sky Survey.  
We therefore expect progress in this direction in the near future. 

\acknowledgments

We would like to thank Michael Blanton, James Bullock, Q. Y. Gong,
Amina Helmi, David Hogg, Savvas Koushiappas, Erin Sheldon, Monica
Valluri, Risa Wechsler, and Beth Willman for useful discussions.  We
are grateful to Frank van den Bosch, Stelios Kazantzidis, Alexander
Knebe, and James Taylor for detailed comments on an earlier draft of
this manuscript and several helpful suggestions.  We thank Tad Pryor
for sharing recent results on the orbital planes of the MW satellites.
We acknowledge use of the publicly-available MW Globular Cluster
Catalog compiled by William Harris.  ARZ is supported by The Kavli
Institute for Cosmological Physics (KICP) at The University of Chicago
and The National Science Foundation (NSF) under grant NSF PHY 0114422.
AVK is supported by the NSF under grants No.  AST-0206216 and
AST-0239759, by NASA through grant NAG5-13274, and by the KICP at the
University of Chicago. OYG is funded by NASA ATP grant NNG04GK68G.
AAK is supported by the NSF grant AST-0206216 to NMSU.  The
simulations presented here were performed on the Origin2000 at the
National Center for Supercomputing Applications (Urbana-Champaign,
Illinois, USA).  This research made use of the NASA Astrophysics Data
System.

\appendix
\section{Satellite Anisotropy In Projection}
\label{app:twod}

Above, we demonstrated that satellite 
halos are {\em not} distributed isotropically about 
their main host halos, but are preferentially aligned 
along the major axes of their hosts.  This, coupled with 
the fact that the luminous dwarfs likely form in a biased 
subset of DM subhalos, greatly mitigates any 
claims that the CDM paradigm of structure formation 
is in conflict with the observed MW satellites.  
Agreement seems to require that the rotation axes of 
disk galaxies are nearly aligned with the major axes of 
their host halos.  
It would be interesting to examine populations 
of satellites around other galaxies and to 
compare them to theoretical predictions.  This 
requires examining the angular distributions 
of satellites in two-dimensional ($2D$) projection, 
where the anisotropy is somewhat diluted.  We present 
the projected distribution of subhalos relative to 
their host halos in this Appendix.

To determine the projected subhalo 
angular distributions, we take 
three orthogonal projections of each 
host halo and find the principal axes of the 
$2D$ mass distributions by diagonalizing 
the $2D$ analog of the moment of inertia 
tensor [Eq.~(\ref{eq:itensor})].  
For each satellite, we determine the angle 
$\phi$, between the major axis of inertia 
and the satellite position.  In practice, 
correlated galaxies along the line-of-sight 
to the host galaxy contribute to the 
differential angular distribution as it is 
difficult to remove objects at a fixed 
three-dimensional distance.  To account 
approximately for this, we include in our 
sample all halos and subhalos within a 
three-dimensional distance $\Rcut = 2 \Rvir$ 
from the host halo center, and within a 
projected, $2D$ distance $\rp = 300 \hkpc$.  
It is not possible to extend our projection to 
larger three-dimensional radius because 
$2 \Rvir$ marks the maximum extent 
of the high-resolution regions of the simulation 
(see \S~\ref{sub:sims}).  However, in a forthcoming 
study we examine the angular distributions 
of halos relative to nearby hosts in projection and 
find that the differential distributions are 
not strongly affected by extending the projection 
region (Zentner et al., in preparation).

\begin{figure}[t]
\epsscale{1.0}
\plotone{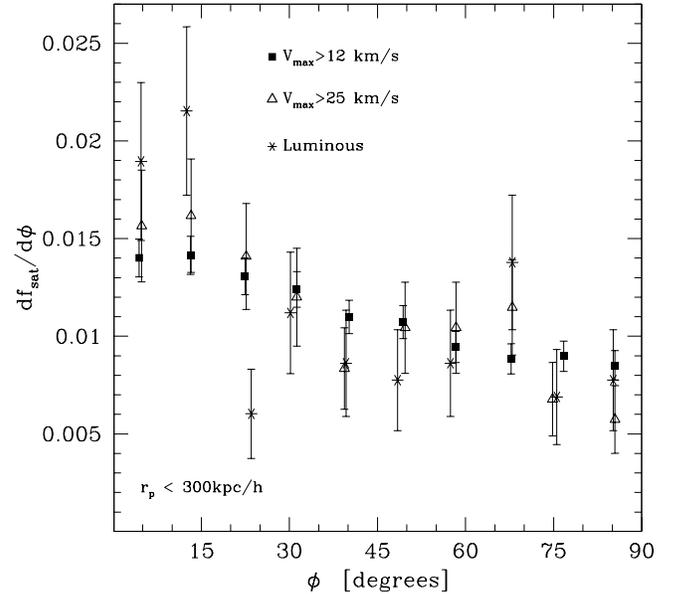}
\caption{
The distribution of subhalo angular position with respect to 
the long axis of the host halo in $2D$ projection.  
The fractional distribution of subhalos $\dd \fsat /\dd \phi$, 
was computed by summing over three 
orthogonal projections for each host halo.  
We show the distribution for all halos and subhalos within 
a three-dimensional distance $\Rcut \le 2 \Rvir$ and a projected 
distance of $\rp \le 300 \hkpc$ of the center of the host halo.
The {\em filled squares} represent all subhalos with 
$\Vmax \ge 12 \kms$.  The {\em open triangles} 
represent all subhalos with $\Vmax \ge 25 \kms$.  
The {\em stars} represent the distribution of 
satellites that should host luminous galaxies 
according to the model of \citet{kravtsov_etal04}.  
}
\label{fig:fsatphi}
\end{figure}

Figure~\ref{fig:fsatphi} shows differential 
distributions of $\phi$ for the satellites 
of our simulated halos.  The anisotropy in 
$2D$ projection remains clear.  
The probability that a satellite lies at the 
major axis ($\phi \le 10\degrees$) 
of the ellipse is $\gsim 50 \%$ larger 
than the probability that it lies near the minor 
axis ($\phi \ge 80\degrees$).  
Additionally, the degree of anisotropy appears 
to be weakly dependent on subhalo size 
over the relatively small range of 
subhalo $\Vmax$ that we can reliably probe.

The coordinate system used in any observational study 
will be defined by the distribution of light rather than 
by the unseen distribution of DM.  As such,  
Fig.~\ref{fig:fsatphi} requires a strong correlation between 
the orientation of the luminous component of the host galaxy 
and the principal axes of its host halo.  In the absence of 
such a correlation, any anisotropy would be diluted away 
in a study of many host systems.  This fact provides an 
interesting constraint on theories of galaxy formation, 
as it can potentially serve to provide some statistical 
measure of the orientations of disk and/or elliptical 
galaxies within their host halos.  

\bibliography{satdir}

\begin{thebibliography}{111}
\expandafter\ifx\csname natexlab\endcsname\relax\def\natexlab#1{#1}\fi

\bibitem[{{Abadi} {et~al.}(2003){Abadi}, {Navarro}, {Steinmetz}, \&
  {Eke}}]{abadi_etal03}
{Abadi}, M.~G., {Navarro}, J.~F., {Steinmetz}, M., \& {Eke}, V.~R. 2003, \apj,
  597, 21

\bibitem[{{Armandroff}(1989)}]{armandroff89}
{Armandroff}, T.~E. 1989, \aj, 97, 375

\bibitem[{{Barnes} \& {Efstathiou}(1987)}]{barnes_efstathiou87}
{Barnes}, J.~A. \& {Efstathiou}, G. 1987, ApJ, 319, 575

\bibitem[{{Barnes} \& {Hernquist}(1996)}]{barnes_hernquist96}
{Barnes}, J.~E. \& {Hernquist}, L. 1996, ApJ, 471, 115

\bibitem[{{Benson} {et~al.}(2004){Benson}, {Lacey}, {Frenk}, {Baugh}, \&
  {Cole}}]{benson_etal04}
{Benson}, A.~J., {Lacey}, C.~G., {Frenk}, C.~S., {Baugh}, C.~M., \& {Cole}, S.
  2004, \mnras, 351, 1215

\bibitem[{{Binney}(1977)}]{binney77}
{Binney}, J.~J. 1977, MNRAS, 181, 735

\bibitem[{{Blumenthal} {et~al.}(1984){Blumenthal}, {Faber}, {Primack}, \&
  {Rees}}]{blumenthal_etal84}
{Blumenthal}, G.~R., {Faber}, S.~M., {Primack}, J.~R., \& {Rees}, M.~J. 1984,
  Nature, 311, 517

\bibitem[{{Bond} {et~al.}(1996){Bond}, {Kofmann}, \& {Pogosyan}}]{bond_etal96}
{Bond}, J.~R., {Kofmann}, L., \& {Pogosyan}, D. 1996, Nature, 380, 603

\bibitem[{{Brainerd}(2004)}]{brainerd04}
{Brainerd}, T.~G. 2004, ApJL, Submitted (astro-ph/0408559)

\bibitem[{{Bullock}(2002)}]{bullock02}
{Bullock}, J.~S. 2002, in in Proceedings of the Yale Cosmology Workshop "The
  Shapes of Galaxies and Their Dark Matter Halos", (28-30 May, 2001).
  P.Natarajan (ed.). Singapore: World Scientific, 109

\bibitem[{{Bullock} {et~al.}(2001){Bullock}, {Kolatt}, {Sigad}, {Somerville},
  {Kravtsov}, {Klypin}, {Primack}, \& {Dekel}}]{bullock_etal01}
{Bullock}, J.~S., {Kolatt}, T.~S., {Sigad}, Y., {Somerville}, R.~S.,
  {Kravtsov}, A.~V., {Klypin}, A.~A., {Primack}, J.~R., \& {Dekel}, A. 2001,
  MNRAS, 321, 559

\bibitem[{{Bullock} {et~al.}(2000){Bullock}, {Kravtsov}, \&
  {Weinberg}}]{bullock_etal00}
{Bullock}, J.~S., {Kravtsov}, A.~V., \& {Weinberg}, D.~H. 2000, \apj, 539, 517

\bibitem[{{Chen} {et~al.}(2003){Chen}, {Jing}, \& {Yoshikaw}}]{chen_etal03}
{Chen}, D.~N., {Jing}, Y.~P., \& {Yoshikaw}, K. 2003, \apj, 597, 35

\bibitem[{{Colberg} {et~al.}(2004){Colberg}, {Krughoff}, \&
  {Connolly}}]{colberg_etal05}
{Colberg}, J.~M., {Krughoff}, K.~S., \& {Connolly}, A.~J. 2004, {\mnras} in
  press (astro-ph/0406665)

\bibitem[{{Col{\'{\i}}n} {et~al.}(1999){Col{\'{\i}}n}, {Klypin}, {Kravtsov}, \&
  {Khokhlov}}]{colin_etal99}
{Col{\'{\i}}n}, P., {Klypin}, A.~A., {Kravtsov}, A.~V., \& {Khokhlov}, A.~M.
  1999, \apj, 523, 32

\bibitem[{{Colless} {et~al.}(2001){Colless}, {Dalton}, {Maddox}, {Sutherland},
  \& {the 2dF collaboration}}]{colless_etal01}
{Colless}, M., {Dalton}, G., {Maddox}, S., {Sutherland}, W., \& {the 2dF
  collaboration}. 2001, \mnras, 328, 1039

\bibitem[{{De Lucia} {et~al.}(2004){De Lucia}, {Kauffmann}, {Springel},
  {White}, {Lanzoni}, {Stoehr}, {Tormen}, \& {Yoshida}}]{delucia_etal04}
{De Lucia}, G., {Kauffmann}, G., {Springel}, V., {White}, S.~D.~M., {Lanzoni},
  B., {Stoehr}, F., {Tormen}, G., \& {Yoshida}, N. 2004, \mnras, 348, 333

\bibitem[{{Diemand} {et~al.}(2004){Diemand}, {Moore}, \&
  {Stadel}}]{diemand_etal04}
{Diemand}, J., {Moore}, B., \& {Stadel}, J. 2004, \mnras, 352, 535

\bibitem[{{Doroshkevich}(1970)}]{doroshkevich70}
{Doroshkevich}, A.~G. 1970, Astrofizika, 6, 581

\bibitem[{{Dubinski}(1994)}]{dubinski94}
{Dubinski}, J. 1994, ApJ, 431, 617

\bibitem[{{Dubinski} \& {Carlberg}(1991)}]{dubinski_carlberg91}
{Dubinski}, J. \& {Carlberg}, R.~G. 1991, \apj, 378, 496

\bibitem[{{Efstathiou} \& {Jones}(1979)}]{efstathiou_jones79}
{Efstathiou}, G. \& {Jones}, B.~J.~T. 1979, MNRAS, 186, 133

\bibitem[{{Fall} \& {Efstathiou}(1980)}]{fall_efstathiou80}
{Fall}, S.~M. \& {Efstathiou}, G. 1980, MNRAS, 193, 189

\bibitem[{{Faltenbacher} {et~al.}(2005){Faltenbacher}, {Allgood}, {Gottloeber},
  {Yepes}, \& {Hoffman}}]{faltenbacher_etal05}
{Faltenbacher}, A., {Allgood}, B., {Gottloeber}, S., {Yepes}, G., \& {Hoffman},
  Y. 2005, {\mnras} submitted (astro-ph/0501452)

\bibitem[{{Font} {et~al.}(2001){Font}, {Navarro}, {Stadel}, \&
  {Quinn}}]{font_etal01}
{Font}, A.~S., {Navarro}, J.~F., {Stadel}, J., \& {Quinn}, T. 2001, \apjl, 563,
  L1

\bibitem[{{Frenk} \& {White}(1982)}]{frenk_white82}
{Frenk}, C.~S. \& {White}, S.~D.~M. 1982, \mnras, 198, 173

\bibitem[{{Gao} {et~al.}(2004{\natexlab{a}}){Gao}, {De Lucia}, {White}, \&
  {Jenkins}}]{gao_etal04b}
{Gao}, L., {De Lucia}, G., {White}, S.~D.~M., \& {Jenkins}, A.
  2004{\natexlab{a}}, \mnras, 352, L1

\bibitem[{{Gao} {et~al.}(2004{\natexlab{b}}){Gao}, {White}, {Jenkins},
  {Stoehr}, \& {Springel}}]{gao_etal04a}
{Gao}, L., {White}, S.~D.~M., {Jenkins}, A., {Stoehr}, F., \& {Springel}, V.
  2004{\natexlab{b}}, MNRAS submitted ({\tt astro-ph/0404589})

\bibitem[{{Gerhard} \& {Binney}(1985)}]{gerhard_binney85}
{Gerhard}, O.~E. \& {Binney}, J.~J. 1985, MNRAS, 216, 467

\bibitem[{{Ghigna} {et~al.}(2000){Ghigna}, {Moore}, {Governato}, {Lake},
  {Quinn}, \& {Stadel}}]{ghigna_etal00}
{Ghigna}, S., {Moore}, B., {Governato}, F., {Lake}, G., {Quinn}, T., \&
  {Stadel}, J. 2000, \apj, 544, 616

\bibitem[{{Ghigna} {et~al.}(1998){Ghigna}, {Moore}, {Governato}, \&
  {Stadel}}]{ghigna_etal98}
{Ghigna}, S., {Moore}, B., {Governato}, F., \& {Stadel}, J. 1998, \mnras, 300,
  146

\bibitem[{{Gnedin} {et~al.}(2004){Gnedin}, {Kravtsov}, {Klypin}, \&
  {Nagai}}]{gnedin_etal04}
{Gnedin}, O.~Y., {Kravtsov}, A.~V., {Klypin}, A.~A., \& {Nagai}, D. 2004, \apj,
  616, 16

\bibitem[{{Grebel} {et~al.}(2003){Grebel}, {Gallagher}, \&
  {Harbeck}}]{grebel_etal03}
{Grebel}, E.~K., {Gallagher}, J.~S., \& {Harbeck}, D. 2003, AJ, 125, 1926

\bibitem[{{Grebel} {et~al.}(1999){Grebel}, {Kolatt}, \&
  {Brandner}}]{grebel_etal99}
{Grebel}, E.~K., {Kolatt}, T., \& {Brandner}, W. 1999, in IAU Symposium, 447

\bibitem[{{Harris}(1996)}]{harris96}
{Harris}, W.~E. 1996, \aj, 112, 1487

\bibitem[{{Hartwick}(1996)}]{hartwick96}
{Hartwick}, F.~D.~A. 1996, in ASP Conf. Ser. 92, Formation of the Galactic
  Halo...Inside and Out, 444

\bibitem[{{Hartwick}(2000)}]{hartwick00}
{Hartwick}, F.~D.~A. 2000, \aj, 119, 2248

\bibitem[{{Hawley} \& {Peebles}(1975)}]{hawley_peebles75}
{Hawley}, D.~L. \& {Peebles}, P.~J.~E. 1975, \aj, 80, 477

\bibitem[{{Hayashi} {et~al.}(2003){Hayashi}, {Navarro}, {Taylor}, {Stadel}, \&
  {Quinn}}]{hayashi_etal03}
{Hayashi}, D., {Navarro}, J.~F., {Taylor}, J.~E., {Stadel}, J., \& {Quinn}, T.
  2003, \apj, 584, 541

\bibitem[{{Helmi}(2004a)}]{helmi04a}
{Helmi}, A. 2004a, \mnras, 351, 643

\bibitem[{{Helmi}(2004b)}]{helmi04b}
---. 2004b, \apjl, 610, L97

\bibitem[{{Helmi} \& {White}(1999)}]{helmi_white99}
{Helmi}, A. \& {White}, S.~D.~M. 1999, \mnras, 307, 495

\bibitem[{{Hoekstra} {et~al.}(2004){Hoekstra}, {Yee}, \&
  {Gladders}}]{hoekstra_etal04}
{Hoekstra}, H., {Yee}, H.~K.~C., \& {Gladders}, M.~D. 2004, \apj, 606, 67

\bibitem[{{Holmberg}(1969)}]{holmberg69}
{Holmberg}, E. 1969, Arkiv Astron., 5, 305

\bibitem[{{Huang} \& {Carlberg}(1997)}]{huang_carlberg97}
{Huang}, S. \& {Carlberg}, R.~G. 1997, ApJ, 480, 503

\bibitem[{{Ibata} {et~al.}(2001){Ibata}, {Lewis}, {Irwin}, {Totten}, \&
  {Quinn}}]{ibata_etal01}
{Ibata}, R., {Lewis}, G.~F., {Irwin}, M., {Totten}, E., \& {Quinn}, T. 2001,
  \apj, 551, 294

\bibitem[{{Ibata} \& {Razoumov}(1998)}]{ibata_razoumov98}
{Ibata}, R.~A. \& {Razoumov}, A.~O. 1998, Astron. \& Astrophys., 336, 130

\bibitem[{{Jing} \& {Suto}(2002)}]{jing_suto02}
{Jing}, Y.~P. \& {Suto}, Y. 2002, \apj, 574, 538

\bibitem[{{Johnston} {et~al.}(2004){Johnston}, {Law}, \&
  {Majewski}}]{johnston_etal04}
{Johnston}, K.~V., {Law}, D.~R., \& {Majewski}, S.~R. 2004, {\apj} submitted
  (astro-ph/0407565)

\bibitem[{{Kang} {et~al.}(2005){Kang}, {Mao}, {Gao}, \& {Jing}}]{kang_etal05}
{Kang}, X., {Mao}, S., {Gao}, L., \& {Jing}, Y.~P. 2005, Astron. \& Astrophys.
  Submitted (astro-ph/0501333)

\bibitem[{{Kauffmann} {et~al.}(1993){Kauffmann}, {White}, \&
  {Guiderdoni}}]{kauffmann_etal93}
{Kauffmann}, G., {White}, S.~D.~M., \& {Guiderdoni}, B. 1993, \mnras, 264, 201

\bibitem[{{Kazantzidis} {et~al.}(2004a){Kazantzidis}, {Kravtsov}, {Zentner},
  {Allgood}, {Nagai}, \& {Moore}}]{kazantzidis_etal04}
{Kazantzidis}, S., {Kravtsov}, A.~V., {Zentner}, A.~R., {Allgood}, B.~A.,
  {Nagai}, D., \& {Moore}, B. 2004a, ApJL, 611, L73

\bibitem[{{Kazantzidis} {et~al.}(2004b){Kazantzidis}, {Mayer}, {Mastropietro},
  {Diemand}, {Stadel}, \& {Moore}}]{kazantzidis_etal04b}
{Kazantzidis}, S., {Mayer}, L., {Mastropietro}, C., {Diemand}, J., {Stadel},
  J., \& {Moore}, B. 2004b, \apj, 608, 663

\bibitem[{{Klypin} {et~al.}(1999b){Klypin}, {Gottl{\"{o}}ber}, Kravtsov, \&
  {Khokhlov}}]{klypin_etal99b}
{Klypin}, A.~A., {Gottl{\"{o}}ber}, S., Kravtsov, A.~V., \& {Khokhlov}, A.~M.
  1999b, \apj, 516, 530

\bibitem[{{Klypin} {et~al.}(2001){Klypin}, {Kravtsov}, {Bullock}, \&
  {Primack}}]{klypin_etal01}
{Klypin}, A.~A., {Kravtsov}, A.~V., {Bullock}, J.~S., \& {Primack}, J.~R. 2001,
  \apj, 554, 903

\bibitem[{{Klypin} {et~al.}(1999a){Klypin}, {Kravtsov}, {Valenzuela}, \&
  {Prada}}]{klypin_etal99a}
{Klypin}, A.~A., {Kravtsov}, A.~V., {Valenzuela}, O., \& {Prada}, F. 1999a,
  \apj, 522, 82

\bibitem[{{Klypin} \& {Shandarin}(1983)}]{klypin_shandarin83}
{Klypin}, A.~A. \& {Shandarin}, S.~F. 1983, MNRAS, 204, 891

\bibitem[{{Knebe} {et~al.}(2004){Knebe}, {Gill}, {Gibson}, {Lewis}, {Ibata}, \&
  {Dopita}}]{knebe_etal04}
{Knebe}, A., {Gill}, S.~P.~D., {Gibson}, B.~K., {Lewis}, G.~F., {Ibata}, R.~A.,
  \& {Dopita}, M.~A. 2004, \apj, 603, 7

\bibitem[{{Kravtsov}(1999)}]{kravtsov99}
{Kravtsov}, A.~V. 1999, PhD thesis, New Mexico State University

\bibitem[{{Kravtsov} {et~al.}(2004){Kravtsov}, {Gnedin}, \&
  {Klypin}}]{kravtsov_etal04}
{Kravtsov}, A.~V., {Gnedin}, O.~Y., \& {Klypin}, A.~A. 2004, \apj, 609, 482
  (KGK04)

\bibitem[{{Kravtsov} {et~al.}(1997){Kravtsov}, {Klypin}, \&
  {Khokhlov}}]{kravtsov_etal97}
{Kravtsov}, A.~V., {Klypin}, A.~A., \& {Khokhlov}, A.~M. 1997, ApJS, 111, 73

\bibitem[{{Kroupa} {et~al.}(2005){Kroupa}, {Theis}, \& {Boily}}]{kroupa_etal04}
{Kroupa}, P., {Theis}, C., \& {Boily}, C.~M. 2005, Astron. \& Astrophys., 431,
  517

\bibitem[{{Law} {et~al.}(2004){Law}, {Majewski}, {Johnston}, \&
  {Strutskie}}]{law_etal04}
{Law}, D.~R., {Majewski}, S.~R., {Johnston}, K.~V., \& {Strutskie}, M.~F. 2004,
  in Satellites and Tidal Streams

\bibitem[{{Lynden-Bell}(1982)}]{lynden-bell82}
{Lynden-Bell}, D. 1982, Obs., 102, 202

\bibitem[{{MacGillivray} {et~al.}(1982){MacGillivray}, {Dodd}, {McNally}, \&
  {Corwin}}]{macgillivray_etal82}
{MacGillivray}, H.~T., {Dodd}, R.~J., {McNally}, B.~V., \& {Corwin}, H.~G.
  1982, \mnras, 198, 605

\bibitem[{{Majewski}(1994)}]{majewski94}
{Majewski}, S.~R. 1994, ApJL, 431, L17

\bibitem[{{Majewski} {et~al.}(2004){Majewski}, {Kunkel}, {Law}, {Patterson},
  {Polak}, {Rocha-Pinto}, {Crane}, {Frinchaboy}, {Hummels}, {Johnston}, {Rhee},
  {Skrutskie}, \& {Weinberg}}]{majewski_etal04}
{Majewski}, S.~R., {Kunkel}, W.~E., {Law}, D.~R., {Patterson}, R.~J., {Polak},
  A.~A., {Rocha-Pinto}, H.~J., {Crane}, J.~D., {Frinchaboy}, P.~M., {Hummels},
  C.~B., {Johnston}, K.~V., {Rhee}, J., {Skrutskie}, M.~F., \& {Weinberg}, M.
  2004, \aj, 128, 245

\bibitem[{{Majewski} {et~al.}(2003){Majewski}, {Skrutskie}, {Weinberg}, \&
  {Ostheimer}}]{majewski_etal03}
{Majewski}, S.~R., {Skrutskie}, M.~F., {Weinberg}, M.~D., \& {Ostheimer}, J.~C.
  2003, \apj, 599, 1082

\bibitem[{{Mart{\'{\i}}nez-Delgado} {et~al.}(2004){Mart{\'{\i}}nez-Delgado},
  {G{\'{o}}mez-Flechoso}, {Aparicio}, \& {Carrera}}]{martinez_delgado_etal04}
{Mart{\'{\i}}nez-Delgado}, D., {G{\'{o}}mez-Flechoso}, M.~{\'{A}}., {Aparicio},
  A., \& {Carrera}, R. 2004, ApJ, 601, 242

\bibitem[{{Mateo}(1998)}]{mateo98}
{Mateo}, M.~L. 1998, \araa, 36, 435

\bibitem[{{Merritt} \& {Quinlan}(1998)}]{merritt_quinlan98}
{Merritt}, D. \& {Quinlan}, G.~D. 1998, ApJ, 498, 625

\bibitem[{{Moore} {et~al.}(1999){Moore}, {Ghigna}, {Governato}, {Lake},
  {Quinn}, {Stadel}, \& {Tozzi}}]{moore_etal99}
{Moore}, B., {Ghigna}, S., {Governato}, F., {Lake}, G., {Quinn}, T., {Stadel},
  J., \& {Tozzi}, P. 1999, ApJL, 524, L9

\bibitem[{{Nagai} \& {Kravtsov}(2005)}]{nagai_kravtsov05}
{Nagai}, D. \& {Kravtsov}, A.~V. 2005, \apj, 618, 557

\bibitem[{{Navarro} {et~al.}(2004){Navarro}, {Abadi}, \&
  {Steinmetz}}]{navarro_etal04}
{Navarro}, J.~F., {Abadi}, M.~G., \& {Steinmetz}, M. 2004, \apjl, 613, L41

\bibitem[{{Navarro} {et~al.}(1994){Navarro}, {Frenk}, \&
  {White}}]{navarro_etal94}
{Navarro}, J.~F., {Frenk}, C.~S., \& {White}, S.~D.~M. 1994, \mnras, 267, L1

\bibitem[{{Navarro} {et~al.}(1995){Navarro}, {Frenk}, \&
  {White}}]{navarro_etal95}
---. 1995, \mnras, 275, 56

\bibitem[{{Navarro} {et~al.}(1997){Navarro}, {Frenk}, \&
  {White}}]{navarro_etal97}
---. 1997, ApJ, 490, 493 (NFW)

\bibitem[{{Pe{\~{n}}arrubia} {et~al.}(2002){Pe{\~{n}}arrubia}, {Kroupa}, \&
  {Boily}}]{penarrubia_etal02}
{Pe{\~{n}}arrubia}, J., {Kroupa}, P., \& {Boily}, C.~M. 2002, \mnras, 333, 779

\bibitem[{{Peebles}(1969)}]{peebles69}
{Peebles}, P.~J.~E. 1969, ApJ, 155, 393

\bibitem[{{Piatek} {et~al.}(2005){Piatek}, {Pryor}, {Bristow}, {Olszewski},
  {Harris}, {Mateo}, {Minniti}, \& {Tinney}}]{piatek_etal05}
{Piatek}, S., {Pryor}, C., {Bristow}, P., {Olszewski}, E.~W., {Harris}, H.~C.,
  {Mateo}, M., {Minniti}, D., \& {Tinney}, C.~G. 2005, AJ, Submitted

\bibitem[{{Porciani} {et~al.}(2002){Porciani}, {Dekel}, \&
  {Hoffmann}}]{porciani_etal02}
{Porciani}, C., {Dekel}, A., \& {Hoffmann}, Y. 2002, MNRAS, 332, 339

\bibitem[{{Quinn} \& {Goodman}(1986)}]{quinn_goodman86}
{Quinn}, P.~J. \& {Goodman}, J. 1986, \apj, 309, 472

\bibitem[{{Reed} {et~al.}(2003){Reed}, {Gardner}, {Quinn}, {Stadel}, {Fardal},
  {Lake}, \& {Governato}}]{reed_etal03}
{Reed}, D., {Gardner}, J., {Quinn}, T., {Stadel}, J., {Fardal}, M., {Lake}, G.,
  \& {Governato}, F. 2003, \mnras, 346, 565

\bibitem[{{Reed} {et~al.}(2004){Reed}, {Governato}, {Quinn}, {Gardner},
  {Stadel}, \& {Lake}}]{reed_etal04}
{Reed}, D., {Governato}, F., {Quinn}, T., {Gardner}, J., {Stadel}, J., \&
  {Lake}, G. 2004, MNRAS submitted ({\tt astro-ph/0406034})

\bibitem[{{Ricotti} \& {Gnedin}(2004)}]{ricotti_gnedin04}
{Ricotti}, M. \& {Gnedin}, N.~Y. 2004, {\apj} submitted (astro-ph/0408563)

\bibitem[{{Sales} \& {Lambas}(2004)}]{sales_lambas04}
{Sales}, L. \& {Lambas}, D.~G. 2004, \mnras, 348, 1236

\bibitem[{{Sharp} {et~al.}(1979){Sharp}, {Lin}, \& {White}}]{sharp_etal79}
{Sharp}, N.~A., {Lin}, D.~N.~C., \& {White}, S.~D.~M. 1979, \mnras, 187, 287

\bibitem[{{Springel} {et~al.}(2001){Springel}, {White}, {Tormen}, \&
  {Kauffmann}}]{springel_etal01}
{Springel}, V., {White}, S.~D.~M., {Tormen}, G., \& {Kauffmann}, G. 2001,
  \mnras, 328, 726

\bibitem[{{Statler}(1987)}]{statler87}
{Statler}, T.~S. 1987, \apj, 321, 113

\bibitem[{{Stoehr} {et~al.}(2003){Stoehr}, {White}, {Springel}, {Tormen}, \&
  {Yoshida}}]{stoehr_etal03}
{Stoehr}, F., {White}, S.~D.~M., {Springel}, V., {Tormen}, G., \& {Yoshida}, N.
  2003, \mnras, 345, 1313

\bibitem[{{Stoehr} {et~al.}(2002){Stoehr}, {White}, {Tormen}, \&
  {Springel}}]{stoehr_etal02}
{Stoehr}, F., {White}, S.~D.~M., {Tormen}, G., \& {Springel}, V. 2002, \mnras,
  335, L84

\bibitem[{{Strauss} {et~al.}(2002){Strauss}, {Weinberg}, {Lupton}, \& {the SDSS
  collaboration}}]{strauss_etal02}
{Strauss}, M.~A., {Weinberg}, D.~H., {Lupton}, R.~H., \& {the SDSS
  collaboration}. 2002, \aj, 124, 1810

\bibitem[{{Taylor} \& {Babul}(2001)}]{taylor_babul01}
{Taylor}, J.~E. \& {Babul}, A. 2001, \apj, 559, 716

\bibitem[{{Taylor} \& {Babul}(2004)}]{taylor_babul04}
---. 2004, \mnras, 348, 811

\bibitem[{{Taylor} {et~al.}(2003){Taylor}, {Silk}, \& {Babul}}]{taylor_etal03}
{Taylor}, J.~E., {Silk}, J., \& {Babul}, A. 2003, in IAU Symp. 220: Dark Matter
  in Galaxies, 118

\bibitem[{{Tormen}(1997)}]{tormen97}
{Tormen}, G. 1997, MNRAS, 290, 211

\bibitem[{{T{\'{o}}th} \& {Ostriker}(1992)}]{toth_ostriker92}
{T{\'{o}}th}, G. \& {Ostriker}, J.~P. 1992, \apj, 389, 5

\bibitem[{{Udry} \& {Martinet}(1994)}]{udry_martinet94}
{Udry}, S. \& {Martinet}, L. 1994, Astron. \& Astrophys., 281, 314

\bibitem[{{Valluri} \& {Merritt}(1998)}]{valluri_merrit98}
{Valluri}, M. \& {Merritt}, D. 1998, ApJ, 506, 686

\bibitem[{{van den Bosch} {et~al.}(2002){van den Bosch}, {Abel}, {Croft},
  {Hernquist}, \& {White}}]{vdb_etal02}
{van den Bosch}, F.~C., {Abel}, T., {Croft}, R.~A.~C., {Hernquist}, L., \&
  {White}, S.~D.~M. 2002, ApJ, 576, 21

\bibitem[{{Vel{\'{a}}zquez} \& {White}(1999)}]{velazquez_white99}
{Vel{\'{a}}zquez}, H. \& {White}, S.~D.~M. 1999, MNRAS, 304, 254

\bibitem[{{Warren} {et~al.}(1992){Warren}, {Quinn}, {Salmon}, \&
  {Zurek}}]{warren_etal92}
{Warren}, M.~S., {Quinn}, P.~J., {Salmon}, J.~K., \& {Zurek}, W.~H. 1992, ApJ,
  340, 771

\bibitem[{{West} \& {Blakeslee}(2000)}]{west_blakeslee00}
{West}, M.~J. \& {Blakeslee}, J.~P. 2000, ApJL, 543, L27

\bibitem[{{Willman} {et~al.}(2004){Willman}, {Governato}, {Dalcanton}, {Reed},
  \& {Quinn}}]{willman_etal04}
{Willman}, B., {Governato}, F., {Dalcanton}, J.~J., {Reed}, D., \& {Quinn}, T.
  2004, \mnras, 353, 639

\bibitem[{{York} {et~al.}(2000){York}, {Adelman}, {Anderson}, {Anderson},
  {Annis}, \& { the SDSS collaboration}}]{york_etal00}
{York}, D.~G., {Adelman}, J., {Anderson}, J.~E., {Anderson}, S.~F., {Annis},
  J., \& { the SDSS collaboration}. 2000, \aj, 120, 1579

\bibitem[{{Zaritsky} \& {Gonzalez}(1999)}]{zaritsky_gonzalez99}
{Zaritsky}, D. \& {Gonzalez}, A.~H. 1999, PASP, 111, 1508

\bibitem[{{Zaritsky} {et~al.}(1997){Zaritsky}, {Smith}, {Frenk}, \&
  {White}}]{zaritsky_etal97}
{Zaritsky}, D., {Smith}, R., {Frenk}, C.~S., \& {White}, S. D.~M. 1997, ApJL,
  478, L53

\bibitem[{{Zaritsky} \& {White}(1994)}]{zaritsky_white94}
{Zaritsky}, D. \& {White}, S.~D.~M. 1994, ApJ, 435, 599

\bibitem[{{Zentner} {et~al.}(2005){Zentner}, {Berlind}, {Bullock}, {Kravtsov},
  \& {Wechsler}}]{zentner_etal05}
{Zentner}, A.~R., {Berlind}, A.~A., {Bullock}, J.~S., {Kravtsov}, A.~V., \&
  {Wechsler}, R.~H. 2005, ApJ, In Press (astro-ph/0411586), 624

\bibitem[{{Zentner} \& {Bullock}(2003)}]{zentner_bullock03}
{Zentner}, A.~R. \& {Bullock}, J.~S. 2003, ApJ, 598, 49

\bibitem[{{Zinn}(1985)}]{zinn85}
{Zinn}, R. 1985, \apj, 293, 424

\end{thebibliography}


\end{document}